\documentclass{jfm}
\usepackage{graphicx}
\usepackage{amsmath}
\usepackage{amssymb}
\usepackage{epstopdf, epsfig}

\shorttitle{Shape and fission instabilities of ferrofluids in non-uniform magnetic fields}
\shortauthor{T. Vieu, C. Walter}

\title{Shape and fission instabilities of ferrofluids in non-uniform magnetic fields} 

\author{Thibault Vieu\aff{1}
  \corresp{\email{thibault.vieu@u-psud.fr}}
  \and
  Cl\'ement Walter\aff{1}
  \corresp{\email{clement.walter@u-psud.fr}}
  \thanks{T.V. and C.W. have equally contributed to this work.}}
\affiliation{\aff{1}Magist\`ere de Physique Fondamentale, Universit\'e Paris-Saclay, B\^at.\,470,\,F-91405 Orsay, France}

\begin{document}

\maketitle

\begin{abstract}
We study static distributions of ferrofluid submitted to non-uniform magnetic fields. We show how the normal-field instability is modified in the presence of a weak magnetic field gradient. Then we consider a ferrofluid droplet and show how the gradient affects its shape. A rich phase transitions phenomenology is found. We also investigate the creation of droplets by successive splits when a magnet is vertically approached from below and derive theoretical expressions which are solved numerically to obtain the number of droplets and their aspect ratio as function of the field configuration. A quantitative comparison is performed with previous experimental results, as well as with our own experiments, and yields good agreement with the theoretical modeling.
\end{abstract}


\section{Introduction}
\label{sect:introduction}
\begin{figure}
  \centerline{\includegraphics{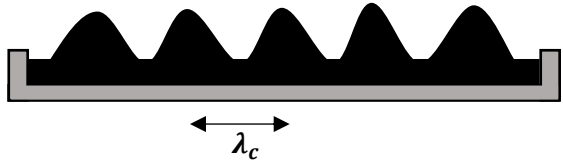}}
  \caption{Normal field instability with  critical wavelength~$\lambda_c$ between the peaks.}
\label{fig:Figure_1}
\end{figure}

Instabilities in ferrofluids under external magnetic fields have been widely studied since the works of \citet{Cowley1967}, from the most fundamental aspects to a wide range of applications. In the presence of a vertical magnetic field, the competition between gravity, surface tension and magnetic energy generates various phenomena, from lattices of spikes in three dimensional layers of fluid, created by the normal-field instability (e.g. \citet{Rosensweig1997,Gailitis1977,Friedrichs2001,Lange2007} and the review by \citet{Odenbach2009}) and characterized by the capillary wavelength $\lambda_c$ (see figure \ref{fig:Figure_1}), to specific patterns in thin ferrofluid layers \citep[e.g.][]{Bourgine2011}, as the labyrinthine patterns discovered by \citet{Cebers1980}. Very thin layers can even break down into isolated peaks thanks to thickness modulation \citep{Petit1993,Bushueva2011} or separated flat droplets \citep{Chen2008}. In the case of the normal-field instability, an other instability is known to furthermore occur through a phase transition between a hexagonal lattice to a square lattice \citep[e.g.][]{Gailitis1977,Abou2000,Gollwitzer2006}.

The behaviour of a single peak created by the normal-field instability has been studied in various field configurations by e.g. \citet{Mahr1998,Friedrichs2000,Lange2000}, and is closely related to the behaviour of a single droplet, which is widely studied since it has promising applications. An important issue motivating the fundamental understanding of droplets manipulation is indeed the creation of microdroplets using non-invasive techniques, as well as their precise control. This leads to direct applications in chemistry, biology and engineering (for a review of the various possible applications of ferrofluids, see \citet{Huebner2008}). Droplets can be created in various ways, e.g. in microfluidics, where flow-focusing geometries can be used to separate droplets and control their size, with or without the help of an external controllable magnetic field, \citep[e.g.][]{Tan2010,Yan2015}.

Droplets dynamics submitted to a lateral uniform field on a hyperhydrophobic surface has also been studied theoretically and experimentally \citep{Brancher1987,Sero1992,Zhu2011}, with the conclusion that the droplet elongates along the direction of the field, which was also evidenced by \citet{Bacri1982} in the case of a droplet immersed in an immiscible fluid. The medium surrounding the droplet can be either magnetic or non magnetic \citep{Rowghanian2016}. The evolution of the droplet's shape is driven by the fluid's surface tension, gravity and the modulation of the demagnetization due to dipole-dipole interactions. The problem is usually treated within a quasi-static approach based on the minimization of the droplet's energy, where its shape is considered to be that of an ellipsoid, as done by e.g. \citet{Bacri1982,Sero1992,Tyler2010}. Such studies often evidence instability phenomena, as the topological instability observed experimentally by \citet{Barkov1980} and \citet{Berkovsky1980}: an isolated droplet of magnetic fluid was observed to fission when the uniform applied magnetic field was increased above some threshold.
This phenomenon was then theoretically described by \citet{Berkovsky1985}
with an approach based on the comparison of the energy of the initial droplet and the energy of the two daughter droplets. The main results of these works are summarized in \citet{Berkovsky1993} with the conclusion that the breakup threshold is more and more difficult to reach as the initial volume is decreased, and the existence of a minimal volume allowing the fission was pointed out. 

\bigbreak

In most of the aforementioned studies, the external magnetic field is uniform and the phenomena mostly driven by the demagnetization dynamics. The effects of a magnetic field gradient have scarcely been considered, although pioneer works should be mentioned, as the theoretical study by \citet{Zelazo1969} where a plane layer of magnetic fluid in a vertical and possibly non-homogeneous magnetic field was considered, and a new dispersion relation obtained. Later, \citet{Catherall2003} experimentally corroborated these findings using paramagnetic liquid oxygen.

Other studies also carried out experiments on various geometries of ferrofluids subject to an inhomogeneous magnetic field in different contexts.
\citet{Rupp2003} built an original experimental setup for their work, where they
submitted a ring of ferrofluid spikes to sinusoidal excitations of varying amplitude, using the inhomogeneous field to trap the droplets along the sharp edge of a cylindrical electromagnet and to induce the excitation. The study of the stable rupture of a ferrofluid layer on a liquid substrate in the shape of a perfect circle when submitted to an inhomogeneous axisymmetric vertically aligned magnetic field was then investigated by \citet{Bushueva2011}, and drops were observed to form periodic well-ordered structures along the boundary of the rupture in the case of sufficiently thick layers. The presence of an inhomogeneous magnetic field was also considered as a stabilizing effect by \citet{Rannacher2006} in their study of solitons on a cylindrical ferrofluid surface, where they showed that a sufficiently strong magnetic field generated by a current-carrying wire at the center of the column of ferrofluid (i.e. a field parallel to the surface) could prevent the breakdown of the fluid column (Rayleigh-plateau instability). Simple models are able to account for such stabilizing effect due to the gradient produced by a current-carrying wire, as reviewed in \citet{Berkovsky1993}. They can also describe the formation of the conical meniscus of fluid surrounding the wire if the latter is initial placed in a pool of ferrofluid \citep{Rosensweig1997}.

\begin{figure}
  \centerline{\includegraphics{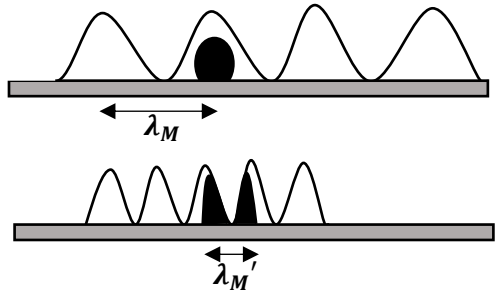}}
  \caption{Illustration of the interpretation from \citet{Timonen2013} of the characteristic magnetic wavelength $\lambda_M$, which decreases when the magnetic field gradient increases (hence when a magnet is approched). When the characteristic wavelength $\lambda_M$ is larger than the droplet's diameter (top panel), the fluid is stable, whereas when the characteristic wavelength $\lambda_M'$ is smaller than the droplet's diameter (bottom), the droplet undergoes a fission.}
\label{fig:Figure_2}
\end{figure}

Recently, \citet{Timonen2013} showed that a ferrofluid droplet submitted to a non-uniform magnetic field is observed to split into two daughter droplets. They mostly focused on describing the dynamical properties of the created lattice of droplets, giving a qualitative explanation of the separation process. They observed that the gravity and the surface tension alone cannot explain the fission of small droplets, and neither can the demagnetization dynamics: a magnetic field gradient seems necessary for this phenomenon to occur. They introduced it through an analogy with the capillary wavelength, providing us with a magnetic wavelength $\lambda_M$, which depends on the field gradient. This critical length is interpreted as follows: a fission is observed when the droplets' diameter is larger than $\lambda_M$ (see figure \ref{fig:Figure_2}).

In this paper, studying a similar system of ferrofluid droplets (of the order of 10 $\umu$L) in a non-uniform magnetic field, we will quantitatively characterize the evolution of the shape of the droplets and derive the number of created droplets as function of the initial volume of fluid, the fluid properties, and the external field configuration. In section \ref{sect:surface_potentials}, we derive with few assumptions a general expression for the potential of a ferrofluid distribution submitted to a non-uniform field. This expression can be used as a basis for further theoretical studies and is suitable for numerical simulations. In section \ref{sect:rosensweig_instability} we show how the peaks of the normal-field instability behave in presence of a weak magnetic field gradient. Then we turn to the main purpose of this paper: the study of the shape and fission of droplets submitted to a non-uniform magnetic field. Sections \ref{sect:aspect_ratio}, \ref{sect:fate} and \ref{sect:fission} provide a theoretical modelling which is solved numerically in section \ref{sect:numerical_resolution} and eventually compared with previous experimental results. Our theoretical results are finally confronted with our experiments in sections \ref{sect:exp_setup} and \ref{results}, focusing on the first separation from one droplet to two daughter droplets.

\section{Theoretical study}
\label{sect:energetic_study}
\begin{figure}
  \centerline{\includegraphics{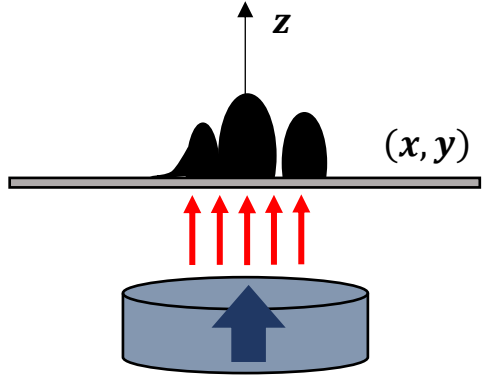}}
\caption{Diagram of the studied system.}
\label{fig:Figure_3}
\end{figure}

Let us consider a distribution of ferrofluid (which may describe a flat layer as well as droplets), lying on a horizontal non-magnetic flat support, vertically aligned with the center of a permanent magnet placed underneath, as illustrated by figure \ref{fig:Figure_3}.

We consider the magnet to be approaching sufficiently slowly to be in the quasi-static regime. From an energetic study, we will derive the distribution of the ferrofluid height $h(x,y)$ on the support.

\subsection{Surface potentials}
\label{sect:surface_potentials}
For a ferrofluid column of height $h(x,y)$ along the $z$ direction and positioned in $(x,y)$, the surface potentials due to gravity and surface tension, and the volume potential due to the magnetic interactions are respectively written \citep{Rosensweig1997,Rosensweig1987,Bourgine2011}:
\begin{gather}
\mathrm{d}^2 U_p[h] = \rho g \int_{0}^{h}  z \, \mathrm{d}z = \rho g \frac{ h^2 }{2} \notag \\
\mathrm{d}^2 U_t[h] = \sigma \sqrt{1+ \left( \boldsymbol{\nabla} h\right) ^2 } \notag \\
\mathrm{d}^3 U_m = - \mu_0 \int_0^{\boldsymbol{H_a}} \mathrm{d}\boldsymbol{h_a} \cdot  \boldsymbol{M(\boldsymbol{H})}
\label{surf_pots}
\end{gather}
with $\rho$ the fluid density, $g$ the gravitational acceleration, $\sigma$ the fluid surface tension coefficient, $\boldsymbol{H_a}$ the applied field, $\boldsymbol{M}$ the ferrofluid magnetization and $\boldsymbol{H}$ the magnetic field modified by the presence of the fluid. In this theoretical study, we neglect the effect of the lying surface, assuming it either superhydrophobic or absent (if for instance the droplet is immersed in another fluid). Note that even if the surface is not superhydrophobic, its contribution to the total energy may be negligible.

For the geometries under consideration (e.g. flat layer or ellipsoidal droplet(s)), the induced field $\boldsymbol{H}$ and the applied field $\boldsymbol{H_a}$ can be related through the following self-consistent equation \citep{Rosensweig1997}:
\begin{equation}
\boldsymbol{H} = \boldsymbol{H_a} - N \boldsymbol{M}(\boldsymbol{H})
\label{self_consistent}
\end{equation}
where $N$ is the demagnetization factor, which encodes the interactions between the magnetic particles inside the fluid. When dealing with a non uniform field, relation (\ref{self_consistent}) is not well-defined. In particular, it is not obvious that a demagnetization factor can be defined. Appendix \ref{appA} shows that in an inhomogeneous field it is not possible to obtain a local relationship of type (\ref{self_consistent}), but it is possible to define a ``demagnetizing series'' $N_n(a;\boldsymbol{r})$, as:
\begin{equation}
H = H_a - \sum_{n=0}^{\infty} N_n(a;\boldsymbol{r}) \left. \frac{\partial^{n} M}{\partial z^{n}} \right|_a
\label{demagnetization_series_intext}
\end{equation}
with $a$ any height inside the fluid distribution.
In this approach, the demagnetization factor $N$ in relation (\ref{self_consistent}) is the first term of the series, $N_0(a;\boldsymbol{r})$. Considering averages, one can show that the second term of the series vanishes, so that relation (\ref{self_consistent}) is also true at first order.

Note that a thorough study of the effect of a non-uniform magnetic field on the magnetization is beyond the scope of our paper. We focus on the effect of the magnetic forces (due to the magnetic field derivatives), and not on the effect of the non-uniform magnetization. This would be an interesting and, as far as we know, new study. However, in most cases, the magnetization curve can be approximated by a linear function (using the differential susceptibility), so that terms beyond the first order in equation (\ref{demagnetization_series_intext}) are not relevant.

We assume the spatial extension of the magnet in the horizontal plane $(x,y)$ to be large compared to that of the ferrofluid. We can thus suppose the magnetic field to be uniform inside the droplets along the horizontal plane, and the magnetization aligned with the external field along the vertical axis. In particular, this amounts to neglecting the horizontal gradient.

For simplicity, we will consider in this section a linear relation for the magnetization:
\begin{gather}
M(H) = \chi H \label{linear_regime_magnetization} \\
H(H_a) = \frac{H_a(z)}{1+ N(z) \chi} 
\label{linear_regime_mag_field}
\end{gather}
and we will reduce the non uniformity of the magnetic field to its gradient:
\begin{equation}
H_a(z) \simeq H_a(0) + z \left. \frac{\partial H_a}{\partial z} \right|_{0}
\label{approx_3}
\end{equation}

A development taking into account the full Taylor expansion in a generalized regime for the magnetization is proposed in Appendix \ref{appB}. Note that in order to compare with experiments we will need to use the generalized regime for the magnetization. The final expression given by equation (\ref{master_equation}) is also suitable for numerical simulations, if one wants for instance to include higher order derivatives of the magnetic field.

In the simplified approach, the magnetic volume potential reads:
\begin{equation}
\mathrm{d}^3 U_m = - \frac{\mu_0}{2} H_a(z)  \chi H(H_a)
\label{mag_surf_pot}
\end{equation}
and the magnetic surface potential is:
%
\begin{multline}
\mathrm{d}^2 U_m[h] =  - \frac{\mu_0}{2} \int_0^h \mathrm{d}z \, H(H_a(0)) \chi  H_a(0)
- \mu_0 \chi H_a(0) \left. \frac{\partial H_a}{\partial z} \right|_{0} \int_0^h \mathrm{d}z \, \frac{z}{1+N(z)\chi} \\
- \frac{\mu_0 \chi}{2} \left( \left. \frac{\partial H_a}{\partial z} \right|_{0} \right)^2 \int_0^h \mathrm{d}z \, \frac{z^2}{1+N(z)\chi} \qquad \qquad \qquad \qquad
\end{multline}

Integrating by parts the second term leads to:
\begin{multline}
\mathrm{d}^2 U_m[h] =  - \frac{\mu_0}{2} \int_0^h \mathrm{d}z \, H(H_a(0)) \chi  H_a(0) + \frac{h^2}{2} F_M
- \frac{\mu_0 \chi}{2} \left( \left. \frac{\partial H_a}{\partial z} \right|_{0} \right)^2 \int_0^h \mathrm{d}z \, \frac{z^2}{1+N(z)\chi}\\
- \mu_0 \chi H_a(0) \left. \frac{\partial H_a}{\partial z} \right|_{0} \int_0^h \mathrm{d}z \, \frac{\chi}{2} \frac{z^2}{(1+N(z)\chi)^2} \frac{\partial N}{\partial z} \qquad \qquad \qquad
\label{surface_pot_simplified}
\end{multline}
where
\begin{equation}
F_M \equiv - \mu_0 \left. \frac{\partial H_a}{\partial z} \right|_{0} \frac{\chi H_a(0)}{1+N(h)\chi}
\end{equation}

The first term of equation (\ref{surface_pot_simplified}) is the usual magnetic potential considered in several papers on the Rosensweig instability, e.g. \citet{Friedrichs2001,Lange2007}. It cannot be directly integrated because $H$ depends on the interface of the fluid distribution through its boundary conditions. 
The second term is the gradient contribution: it comes from the homogeneous volume force applied in the whole ferrofluid distribution by the magnet.
The third term is a squared gradient contribution, which is opposed to the gradient contribution.
The last term is the contribution from the inhomogeneity of the magnetization. Appendix \ref{appA} shows that in most cases this inhomogeneity is not expected to come from the inhomogeneity of the applied field: its main origin should be the geometry of the sample, as for uniformly magnetized samples.

The total surface potential reads:
\begin{multline}
\qquad \quad \mathrm{d}^2 U[h] = \frac{ h^2 }{2} \left( \rho g + F_M \right) 
+ \sigma \sqrt{1+ \left( \boldsymbol{\nabla} h\right) ^2 }
- \frac{\mu_0}{2} \int_0^h \mathrm{d}z \, H(H_a(0)) \chi  H_a(0) \\
- \mu_0 \chi H_a(0) \left. \frac{\partial H_a}{\partial z} \right|_{0} \int_0^h \mathrm{d}z \, \frac{\chi}{2} \frac{z^2}{(1+N(z)\chi)^2} \frac{\partial N}{\partial z}
- \frac{\mu_0 \chi}{2} \left( \left. \frac{\partial H_a}{\partial z} \right|_{0} \right)^2 \int_0^h \mathrm{d}z \, \frac{z^2}{1+N(z)\chi}
\label{linear_m_pot}
\end{multline}
We can already predict from the above expression three different regimes for the external parameters, leading to different behaviours of the ferrofluid:

First, in the limit case of a weak gradient and a high demagnetization, the physics of the ferrofluid is driven by its demagnetization, which corresponds to the study of Rosensweig, and spawns a normal-field instability. We can then keep the leading order terms only, i.e the first 3 terms, and extract a critical magnetic field at which the instability appears, characterised by a modified field-dependent wavelength $\lambda$. We investigate this in section \ref{sect:rosensweig_instability}.

Second, in the high gradient and low demagnetization regime, the gradient plays the main role, essentially through the term $F_M$. 
In this regime, a single droplet can be observed to fission. We focus on this phenomenon in section \ref{sect:fission}.

Third, in the ``in-between'' regime, there exist various instabilities leading to first or second order phase transitions in the evolution of a droplet's aspect ratio. This is described in sections \ref{sect:aspect_ratio} and \ref{sect:fate}.

\begin{figure}
  \centerline{\includegraphics[width=\linewidth]{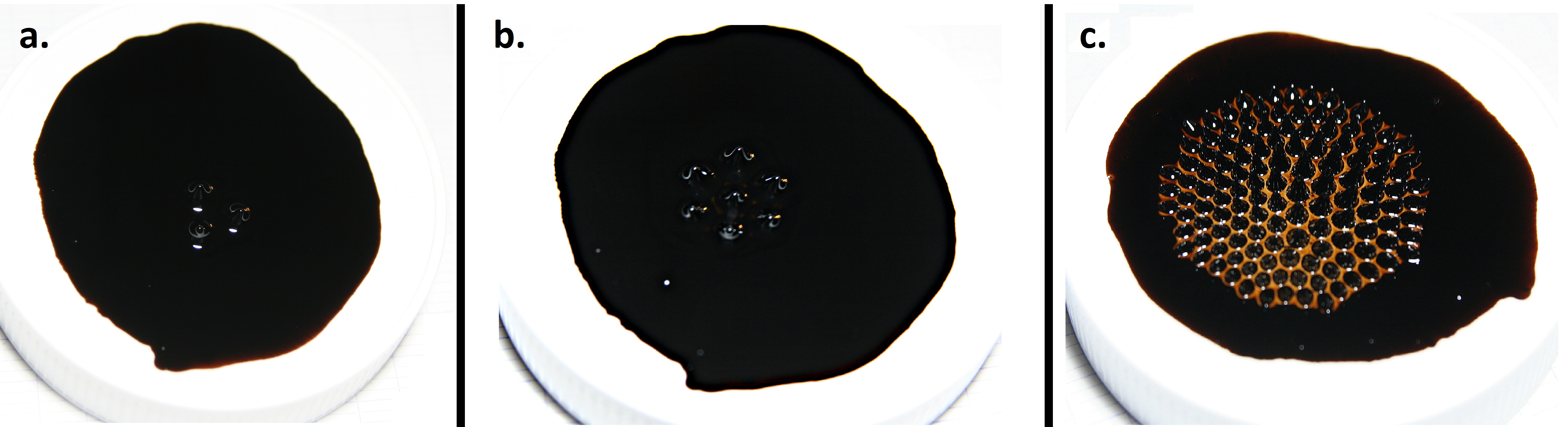}}
\caption{Rosensweig instability experiment in the case of a non-uniform magnetic field. A large neodymium permanent magnet is slowly approached toward a thin layer of ferrofluid (with depth of a few tenth of millimeters) from below. We see in picture (a.) the apparition of the spikes characteristic of the normal-field instability, located above the center of the magnet. The number of spikes increases as the magnet gets closer, as shown in picture (b.). In (c.) we observe that, because of the growing magnetic field gradient, the spikes are getting closer to one another. The spikes are then separated in individual droplets since the original ferrofluid layer is significantly thin.}
\label{fig:Figure_4}
\end{figure}

\subsection{Rosensweig instability in a non-uniform magnetic field}
\label{sect:rosensweig_instability}

The first observed behaviour of the ferrofluid distribution when slowly increasing the magnetic field is the so-called Rosensweig instability. From the more general expression of the surface potential given by equation (\ref{master_equation}), taking the demagnetization to be homogeneous inside the fluid and equal to the one in the usual case of an uniformly magnetized layer, as prescribed by our generalization of the demagnetizing field given in appendix \ref{appA}, and keeping only the first order terms in $h$, we can write the surface potential as:
\begin{equation}
\mathrm{d}^2 U[h] = \frac{h^2}{2} \alpha
+ \sigma \sqrt{1+ \left( \boldsymbol{\nabla} h\right) ^2 }
- \mu_0 \int_0^h \mathrm{d}z \int_0^{H_a(0)} \mathrm{d} h_a \, M(H)
\label{Rosensweig_pot}
\end{equation}
where
\begin{equation}
\alpha \equiv \rho g + \tilde{F_M} \equiv \rho g - \mu_0 \left. \frac{\partial H_a}{\partial z} \right|_{0}
\frac{M(H_a(0))}{1 + N \tilde{\chi}}
\end{equation}
and $\tilde{\chi}$ is the differential susceptibility:
\begin{equation}
\tilde{\chi} \equiv \left. \frac{\partial M}{\partial H} \right|_{H_a(0)}
\end{equation}
%



Equation (\ref{Rosensweig_pot}) is mathematically equivalent to the surface potential usually written when describing the Rosensweig instability of a ferrofluid submitted to an uniform field $H_a(0)$ within an energetic formalism first used by \citet{Gailitis1977}. The non-uniformity of the field reduces, under our assumptions, to a modified contribution of the gravity within~$\alpha$.

From here, writing the height $h$ as a superposition of plane waves, it can be shown that the amplitude $A$ of a perturbation at the surface of the fluid grows as:
\begin{equation}
A \sim \Re \left( e^{-\mathrm{i} \omega t}\right)
\end{equation}
with the frequency $\omega$ related to the wavevector $k$ through the following dispersion relation \citep[e.g.][]{Cowley1967,Rosensweig1997,Salin1993,abou1998,Andelman2009}:
\begin{equation}
\rho \omega^2 = k \left( \sigma k^2 + \alpha - \frac{\mu_0 M^2}{1 + \frac{1}{r}} k \right)
\label{dispersion_relation}
\end{equation}
with $M \equiv M(H(H_a(0)))$ the magnetization, $r$ the dimensionless permeability ratio, $\mu_c$ the chord permeability, $\mu_t$ the tangent permeability:
\begin{equation}
r = \sqrt{\frac{\mu_0^2}{\mu_c \mu_t}} \qquad
\mu_c =  \frac{ \mu_0 H_a(0)}{H(H_a(0))} \qquad
\mu_t =  \mu_0  \left. \frac{\partial H_a}{\partial H} \right|_{H(H_a(0))}
\end{equation}
This dispersion relation is valid without assuming a linear relation for the magnetization \citep{Cowley1967,Andelman2009}. Even though the magnetization is often considered linear, the case of a non-linear magnetization has also been studied by e.g. \citet{Knieling2007,Lange2016}.

If $\omega^2 > 0$, any disturbance will develop at the interface of the fluid, which signals an instability resulting in the creation of a pattern of spikes. The neutral stability is therefore defined by $\omega = 0$, and the flat layer is unstable if there exist a wavevector fulfilling this condition, i.e. the discriminant of equation (\ref{dispersion_relation}) is positive:
\begin{equation}
\left( \frac{\mu_0 M^2}{1+\frac{1}{r}} \right)^2 - 4 \alpha \sigma > 0
\label{instability_condition}
\end{equation}
This defines the critical magnetization (reached at a critical field $H_a^c$), above which the normal-field instability is observed:
\begin{equation}
M^c = \left( 4 \frac{\sigma \alpha}{\mu_0^2 } \left( 1 + \frac{1}{r} \right)^2 \right)^{1/4}
\end{equation}
For $M > M^c$, a pattern of spikes develops at the surface of the fluid. The distance between two spikes is related to the wavelength of the most unstable mode $\omega_m$, i.e. the one with the maximum growth rate \citep[e.g.][]{Bashtovoi1985}:
\begin{equation}
\left. \frac{\mathrm{d}\omega}{\mathrm{d}k} \right|_{\omega_m} = 0
\label{max_growth}
\end{equation}
Together with the dispersion relation (\ref{dispersion_relation}), equation (\ref{max_growth}) gives the characteristic wavevector $k_0$ and the corresponding characteristic wavelength $\lambda$ of the pattern. Placing ourselves at the point of appearance of the instability, we get:
\begin{gather}
k_0 = \sqrt{\alpha/\sigma} \\
\lambda = 2 \upi \sqrt{\frac{\sigma}{\alpha}} = \left( \left( \frac{1}{\lambda_c} \right)^2 + \left( \frac{1}{\lambda_M} \right)^2 \right)^{-1/2}
\label{equation_lambda}
\end{gather}
where we introduced the capillary wavelength $\lambda_c$ (which is the characteristic wavelength of the normal-field instability in uniform field \citep{Rosensweig1997}) and the characteristic wavelength from the magnetic field gradient $\lambda_M$, which are defined by:
\begin{align}
\lambda_c &\equiv 2 \upi \sqrt{\frac{\sigma}{\rho g}}\notag \\
\lambda_M &\equiv 2 \upi \sqrt{\frac{\sigma}{\tilde{F_M}}}
\label{def_characteristic_wl}
\end{align}


These expressions reflect the competition between gravity and surface tension, which tend to flatten the distribution, and the magnetic interactions, which tend to increase the interface area. From equation (\ref{equation_lambda}) we can distinguish two regimes: a low gradient regime, providing us with $\lambda \sim \lambda_c$, in which the uniform contribution is prevalent and the gradient only slightly increases the value of the critical magnetization, and a high gradient regime, in which $\lambda \sim \lambda_M$, dominated by the gradient. If the gradient is too high the instability might never appear (in particular if the critical magnetization is above the saturated magnetization). In this second regime, the critical wavelength has been identified in \citet{Timonen2013} as an important criterion in the description of the fission of ferrofluid droplets.

\begin{figure}
  \centerline{\includegraphics[width=\linewidth]{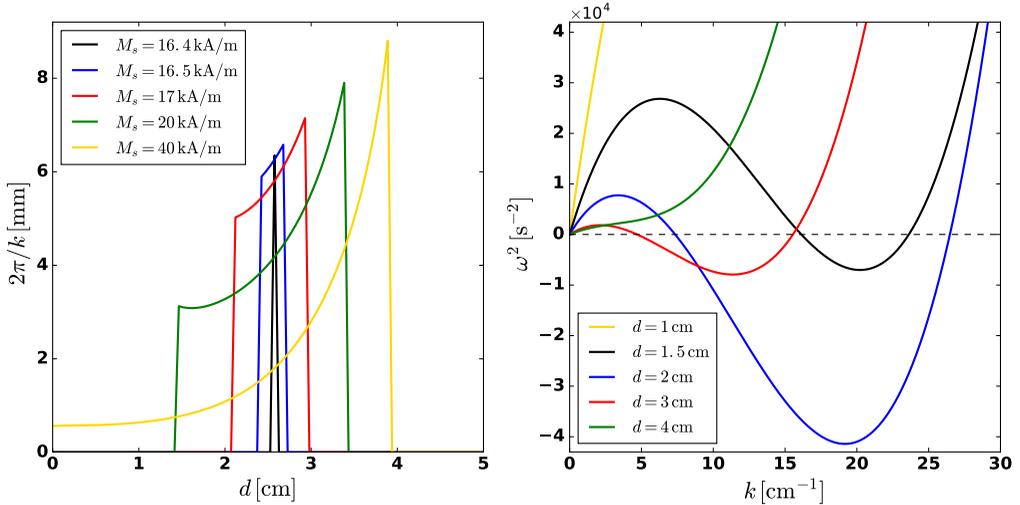}}
  \caption{Left: Evolution of the field-dependent wavelength as function of the distance from a typical neodymium magnet of intrinsic magnetization $M_0 = 950$ kA/m, height 4 cm, radius 1 cm, and various saturated magnetizations. Equation (\ref{magnetic_field_Camacho}) is used to obtain the field strength and field gradient as function of the distance. The other parameters are $\sigma = 50$ mN/m, $\rho = 2000$ kg/m$^3$, $\chi = 1$. Right: Dispersion relation corresponding to $M_s = 20$ kA/m (green curve of the left plot), at different distances from the magnet.}
\label{fig:Figure_5}
\end{figure}

The dispersion relation (\ref{dispersion_relation}) has been obtained by mapping the problem of a non-uniformly magnetized ferrofluid layer to the problem of an uniformly magnetized layer, and applying the linear stability analysis known in the latter case. Interestingly, the full linear stability analysis in the case of an inhomogeneous applied field has been carried out by \citet{Zelazo1969}, and leads to the same dispersion relation. This theoretical result has later been experimentally corroborated by \citet{Catherall2003} using liquid oxygen in a magnetic field gradient. They measured the critical magnetic field and the peak separation as function of an effective gravity $\tilde{g}$ defined, in our notations, as:
\begin{equation}
\tilde{g} \equiv \alpha / \rho
\end{equation}
The relations $M^c \propto \tilde{g}^{1/4}$ and $\lambda \propto \tilde{g}^{-1/2}$ were in very good agreement with the experiments, which shows that the main impact of a magnetic field gradient on a paramagnetic fluid is to redefine the gravitational acceleration. This is also summarized by \citet{Berkovsky1993}, who substitute the gravitational force by an effective value:
\begin{equation}
(\rho g)_{eff} \equiv \rho g - \mu_0 M \frac{\partial H}{\partial z}
\end{equation}
Interestingly, such mathematical analogy is used in \citet{Timonen2013} to describe the fission of a ferrofluid droplet in a magnetic field gradient with the magnetic wavelength $\lambda_M$, although they give no further justification to it. We will investigate this in section~\ref{sect:fission}.

In the case of a layer of fluid, we see that the gradient introduces two important corrections. First the critical magnetization $M^c$ is field-dependent and increases with the gradient: the gradient stabilizes the distribution. Stabilizing effects of non-uniform magnetic fields were indeed observed by \citet{Zelazo1969} and are known in the case of a ferrofluid column surrounding a current-carrying wire \citep[e.g.][]{Rannacher2006}.
Second, the characteristic wavelength $\lambda$ is also field-dependent, so that the distance between the spikes at onset of the instability is impacted by the gradient, in particular not expected to be equal to the capillary wavelength as shown by \citet{Catherall2003}. 

Moreover, pursuing the linear analysis beyond the threshold in the thick-film regime allows one to find the characteristic wavelength of the evolved pattern, corresponding to the wavevector of the most unstable mode, i.e. the solution of equation (\ref{max_growth}):
\begin{equation}
k = \left( \frac{\mu_0 M^2}{1+\frac{1}{r}} + \sqrt{\Delta}\right) \frac{1}{3 \sigma}, \quad \Delta = \left(\frac{\mu_0 M^2}{1+\frac{1}{r}}\right)^2 - 3 \alpha \sigma
\label{k_evolved_pattern}
\end{equation}
which implies that the peaks should get closer and closer as the magnetic field is increased. This result was already obtained by \citet{Abou1997,Lange2000mhd} in the case of an uniform magnetic field. Equation (\ref{k_evolved_pattern}) shows that the increasing gradient of a non-uniform magnetic field slows down this effect. Note that the evolved pattern, made of sharped spikes, is nonlinear so that a quantitative agreement between equation (\ref{k_evolved_pattern}) and experiments is not expected. In order to do relevant quantitative predictions, one has to carry a nonlinear analysis, as done in e.g. \citet{Gailitis1977,Friedrichs2001,Lange2007}. This is left for future work.
We nevertheless investigated this behaviour qualitatively by performing an elementary experiment, considering a thin layer of ferrofluid and slowly increasing the magnetic force by approaching a magnet from below. We observed, beyond a threshold, the appearance of spikes in a hexagonal pattern similar to the one typically observed beyond the normal-field instability threshold, as shown in figure \ref{fig:Figure_4}. Then, when the magnet is approached beyond this threshold, the peaks are observed to get closer and closer, as predicted by equation (\ref{k_evolved_pattern}).

%

Importantly, since the critical magnetization increases with the gradient, there is no guarantee that the condition (\ref{instability_condition}) is always fulfilled after the instability threshold if one keep increasing the field beyond the critical value. A first order phase transition may occur at some point, making the whole distribution to suddenly collapse to a flat interface. This is shown in figure \ref{fig:Figure_5} where is plotted the value of the wavelength as a function of the distance to a typical neodymium magnet (detailed input values are given in the caption), where $1/r$ was put to 1 for simplicity and the magnetic field is computed using the expression of the field created by a finite solenoid \citep[e.g.][]{Camacho2013}:
\begin{equation*}
H_a = \frac{M_0}{2} \left( \frac{d+h}{\sqrt{ (d+h)^{2}+R^2 }} - \frac{d}{\sqrt{ d^{2}+R^2}} \right)
\label{magnetic_field_Camacho}
\end{equation*}

As the magnet approaches (from right to left in the plot), we can first see the appearance of the instability, i.e. the formation of spikes, followed by a smooth but quick decrease of the wavelength, i.e. the decrease of the distance between the spikes. Then, as the contribution of the gradient gets higher, the magnetization may return below the threshold so there is no more instability (the frequency $\omega$ is imaginary). The perturbed interface therefore collapses to the initial flat distribution. Note that in this theoretical example, for $M_s \lesssim 16.4$ kA/m, the normal-field instability never appears.

Such transition was not observable in our qualitative experiment shown in figure \ref{fig:Figure_4} because the magnetization of our ferrofluid was too high, and neither was the morphological transition from the hexagonal pattern to a square pattern.

The kind of phase transition evidenced here, due to the competition between the uniform magnetic field contribution and the magnetic field gradient, will have its equivalent in the case of a single ferrofluid droplet studied in section \ref{sect:aspect_ratio}. The uniform contribution, which elongates the spikes, will equivalently stretch the droplet, while the gradient contribution, which flattens the spikes, will equivalently flatten the droplet until inducing a first order phase transition from a prolate ellipsoid to an oblate ellipsoid. The fundamental mechanism of such transition is exactly the same as for the transition evidenced in figure \ref{fig:Figure_5}.

Just as this last remark shows a bridge between the peaks creation and the evolution of the shape of a droplet, a similar connection can be made with the droplet's fission phenome\-non. Indeed, our qualitative experiment displayed the fission of the peaks once the magnet was close enough to the ferrofluid, like if they were droplets lying on the ferrofluid film they originated from. This phenomenon is similar to the one evidenced by \citet{Timonen2013} for a single droplet on a hyperhydrophobic surface, and on which we focus in section \ref{sect:fission}.
%
We can also see on picture (c.) of figure \ref{fig:Figure_4} that, starting with a very thin layer, we end up at high field with separated spikes which form individual droplets. This change of topology, recently studied by \citet{Bushueva2011}, is due to the fact that the layer is not thick enough for peaks to appear, hence several separated droplets are created instead of several peaks. \citet{Bushueva2011} indeed observed that the critical magnetic field imposing such breakup is proportional to the thickness of the layer.

\subsection{Evolution of the shape of a single ellipsoidal droplet}
\label{sect:aspect_ratio}
We now turn to the main study of this paper and investigate the behaviour of a ferrofluid droplet in a non-uniform magnetic field, again created by approaching a magnet. We do not consider a layer of fluid anymore but an ellipsoidal droplet. The theoretical modeling is in the same spirit as the energetic study done by \citet{Bacri1982}.
\begin{figure}
  \centerline{\includegraphics[width=\textwidth]{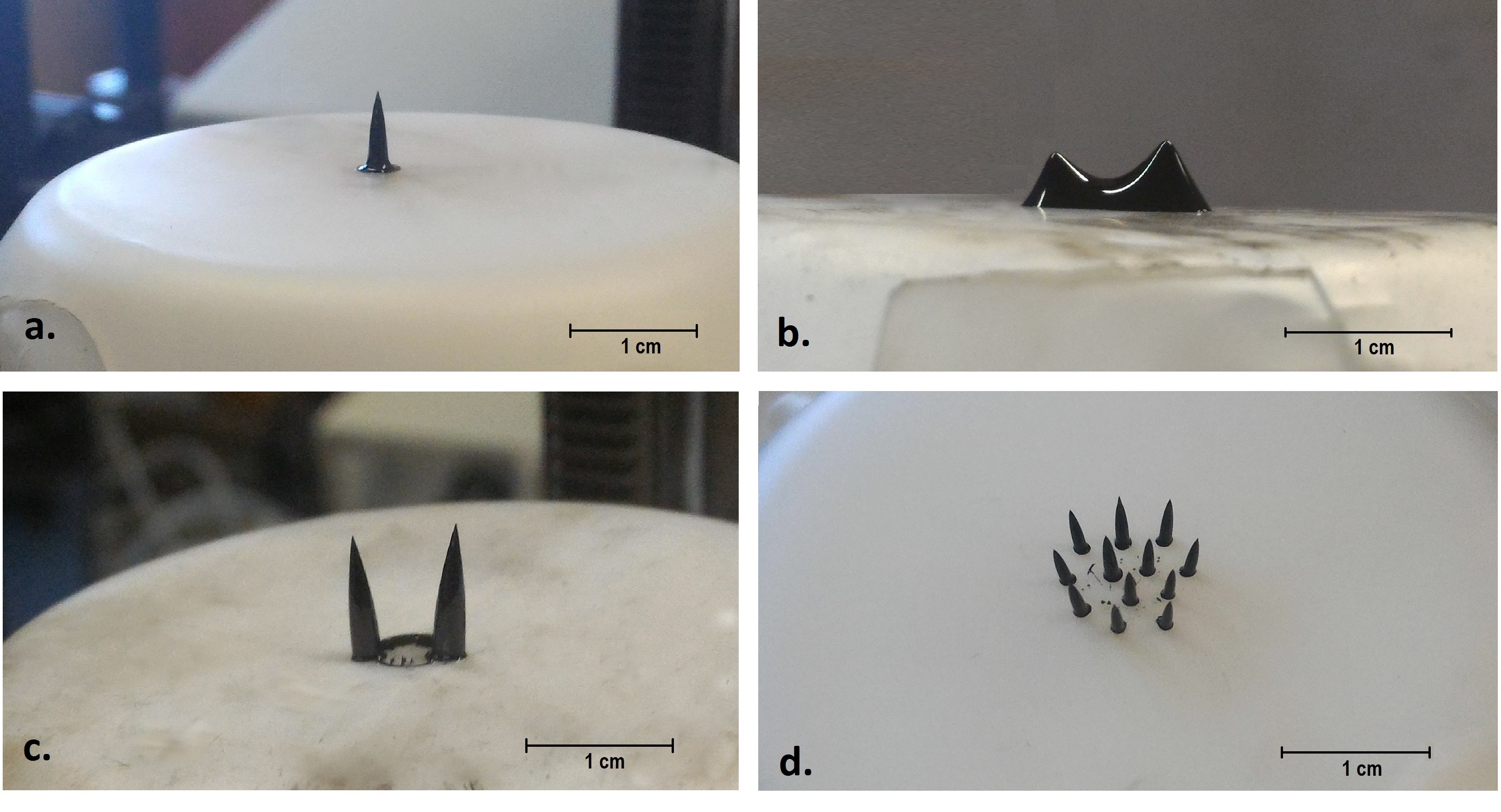}}
  \caption{Typical distributions adopted by the ferrofluid (various initial volumes and external fields), showing the different phases leading to the fissioning of a droplet into several sub-droplets as the applied magnetic field and magnetic field gradient are increased.
a. Vertical stretching. b. Splitting phase. c. Distribution of fluid after fission. d. Creation of several droplets. The amount of fluid is huge in the second picture in order to capture the transition, but it is not representative of our experiments.}
\label{fig:Figure_6}
\end{figure}

The experimental observation we want to explain is the following: when the magnet is approached, moving along the vertical direction, the droplet lengthens and slims until it suddenly splits into smaller droplets. If we bring the magnet even closer, the droplets split again, and so on, creating henceforth numerous droplets, whose volume shrinks with each division. As observed by \citet{Timonen2013}, the splitting is irreversible since the droplets' inertia spawns a distance of separation which can no longer be overcome. Figure \ref{fig:Figure_6} illustrates the steps leading to the fission in several droplets. \citet{Timonen2013} provide high-quality videos of the phenomenon under interest (see in particular movies S1 and S3).

The physical reason underlying this phenomenon is similar to the one responsible for the appearance of the peaks of the normal-field instability: beyond the threshold, the cost of energy of the increased surface of the droplets is smaller than the loss of magnetic energy due to their increased interface.

In this section we focus on the evolution of the aspect ratio, i.e. before the fission occurs. In the next section we investigate the fission phenomenon.

We start the modeling by computing the previously derived potential for an ellipsoidal geometry. In most of the previous studies, the linear regime $M=\chi H$ was assumed for the magnetization. In our case (and also for the setup of \citet{Timonen2013}), we are working with high magnetic fields and the magnetization is in an intermediary regime, closer to saturation. We should therefore use the generalized potential, given in Appendix \ref{appB} by equation (\ref{master_equation}).
We keep the first and second magnetic field derivatives, as well as the gradient squared term, but not the product of the gradient and the second derivative nor higher order terms. As justified in appendix \ref{appA}, we take the demagnetization factor equal to the demagnetization factor of an uniformly magnetized ellipsoid, which is \citep{Bacri1982,Brancher1987,Tyler2010}:
\begin{equation}
N = \frac{1-e^2}{2e^3} \left( -2e+\ln\left( \frac{1+e}{1-e} \right) \right)
\label{averaged_N}
\end{equation}
where $e \equiv \sqrt{1 - K^2}$ is the eccentricity, $K \equiv L/A$ is the inverse aspect ratio, $A$ and $L$ are respectively the major and minor axes.

Under those approximations, the general potential of the ellipsoidal fluid distribution reduces to:
\begin{equation}
U = \alpha \int_0^A \mathrm{d}z \, \mathcal{A}(z) z
+ \sigma S
- \mu_0 V \int_0^{H_a(0)}\mathrm{d}h_a M(H)
- \sigma \frac{2  B_\nabla}{4} \frac{1}{1 + N \tilde{\chi}} \int_0^A \mathrm{d}z \, \mathcal{A}(z) z^2
\label{ellipsoidal_droplet_pot}
\end{equation}
where $\alpha = \rho g + \tilde{F_M}$ and:
\begin{align}
B_\nabla \equiv \frac{\mu_0}{\sigma} \left(
\left. \frac{\partial^2 H_a}{\partial z^2} \right|_{0} M(H_a(0))
+ \left( \left. \frac{\partial H_a}{\partial z} \right|_{0} \right)^2 \tilde{\chi} \right)
\end{align}

$V = \upi A L^2/6$ is the ellipsoid volume,
$\mathcal{A}(z) = L^2 \upi  (A-z) z/A^2$ is the area of the section of the ellipsoid at height~$z$, 
$S= \frac{AL\upi}{2} \left(K + \frac{1}{e}\arcsin e \right)$ is the ellipsoid surface with $e$ its eccentricity.

$A$ and $L$ are related to the inverse aspect ratio thanks to the following:
\begin{equation}
A = 2 R_0 K^{-2/3} \qquad L = 2 R_0 K^{1/3}
\label{geometrical_relations}
\end{equation}
where $R_0$ defined through $V \equiv 4/3 \upi R_0^3$ is the radius of the equivalent sphere.
%
%

Finally the potential $\tilde{U} \equiv U/(2 \upi \sigma)$ is:
\begin{equation}
\tilde{U} = \frac{8 \upi^2 R_0^4}{3} \frac{K^{-2/3}}{\lambda^2}
+ R_0^2 f(K)
- \frac{2 R_0^3}{3} \frac{\mu_0}{\sigma} \int_0^{H_a(0)}\mathrm{d}h_a M(H)
- \frac{2 R_0^5}{5} \frac{B_\nabla K^{-4/3}}{1 + N \tilde{\chi}}
\label{potential_aspect_ratio_1}
\end{equation}
where $\lambda = 2 \upi \sqrt{\frac{\sigma}{\alpha}}$, given by equation (\ref{equation_lambda}), and
\begin{equation}
f(K) \equiv K^{-1/3} \left(K + \frac{\arcsin e}{e} \right)
\label{f_geometrical_function}
\end{equation}

In order to compute the magnetic integral in equation (\ref{potential_aspect_ratio_1}), we use the same approximation as in Appendix \ref{appB}, generalizing the linear regime for the magnetization as follows:
\begin{gather}
M(H) = M(H_a - N M(H)) \simeq M(H_a) - N M(H) \left. \frac{\partial M}{\partial H} \right|_{H_a} \notag \\
M(H) \simeq M(H_a) \left( 1 + N \left. \frac{\partial M}{\partial H} \right|_{H_a} \right)^{-1}
\end{gather}
which is valid either for a small demagnetization factor (which is in general the case for an ellipsoid, $N \lesssim 0.3$) or when the second derivative of the magnetization is negligible (it is in particular exact in the linear regime).

We separate the magnetic integral into two linear parts, defining $H_*$ such that:
\begin{equation}
\chi H_* = M(H_a(0)) - \tilde{\chi} H_a(0) + \tilde{\chi} H_*
\end{equation}
where $\chi$ is the initial susceptibility and $\tilde{\chi} \equiv \left. \frac{\partial M}{\partial H} \right|_{H_a(0)}$ the differential susceptibility. 

The magnetization curve is therefore approximated by two connected straight lines:
\begin{align}
&\forall h_a < H_* \, \, M(h_a) = \chi h_a \notag \\
&\forall h_a > H_* \, \, M(h_a) = M(H_a(0)) - \tilde{\chi} H_a(0) + \tilde{\chi} h_a
\end{align}

The integral becomes:
\begin{align}
\int_0^{H_a(0)}\mathrm{d}h_a M(H) &= \int_0^{H_*}\mathrm{d}h_a M(H) + \int_{H_*}^{H_a(0)}\mathrm{d}h_a M(H)
&= \frac{\sigma}{2 \mu_0} \left( \frac{B_{m,1}}{1 + N \chi} + \frac{B_{m,2}}{1 + N \tilde{\chi}} \right)
\end{align}
where
\begin{align}
B_{m,1} &\equiv \frac{\mu_0}{\sigma} \chi H_*^2  \\
B_{m,2} &\equiv \frac{\mu_0}{\sigma} \tilde{\chi} \left( (H_a(0)^2-H_*^2) + 2 \frac{\chi - \tilde{\chi}}{\tilde{\chi}} H_* (H_a(0) - H_*) \right)
\end{align}

Finally the potential is:
\begin{equation}
\frac{\tilde{U}(K)}{R_0^2} = \frac{3}{2} R_0^2 B_\lambda K^{-2/3}
+ f(K)
- \frac{2 R_0^3}{5} \frac{B_\nabla K^{-4/3}}{1 + N(K) \tilde{\chi}}
- \frac{R_0}{3} \left( \frac{B_{m,1}}{1 + N(K) \chi} 
+ \frac{B_{m,2}}{1 + N(K) \tilde{\chi}} \right)
\label{potential_aspect_ratio_2}
\end{equation}
where
\begin{equation}
B_\lambda \equiv \frac{16 \upi^2}{9 \lambda^2}
\end{equation}
Note that $B_\lambda$ is shape-dependent because of the presence of $\tilde{\chi} N$ in $\tilde{F_M}$. However, this product is very small since we are usually close to the saturated regime. We therefore neglect this shape-dependency in the analytical study. 
\begin{figure}
  \centerline{\includegraphics[width=0.7\linewidth]{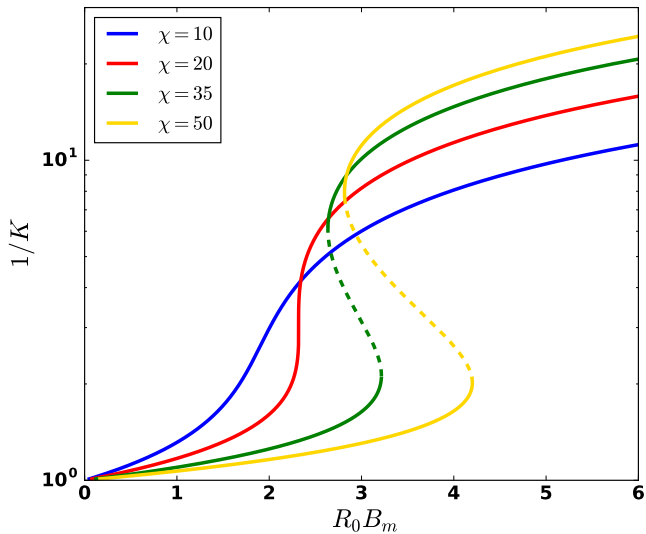}}
\caption{Aspect ratio instability for $B_\lambda = 0$. Solid lines show stable minima, dotted lines show unstable maxima. $R_0 B_m$ is used as control parameter.}
\label{fig:Figure_7}
\end{figure}

This potential, function of four dimensionless control parameters $R_0^2 B_\lambda$, $R_0^3 B_\nabla$, $R_0 B_{m,1}$, $R_0 B_{m,2}$ contains a rich structure depending on the relative strength of the parameters. The evolution of the droplet's shape is driven by the competition between the gradient, the gravity, and the surface tension contributions, which tend to flatten the droplet, and the squared gradient, the second derivative of the magnetic field, and the demagnetization, which tend to stretch the droplet. For simplicity, we will now restrict the analysis to the ``weak gradient regime'' defined by $R_0 B_\nabla \ll B_\lambda$. We therefore fix $B_\nabla$ to zero. Qualitatively, $B_\nabla$ enhances the effect of the demagnetization: $K$ decreases when $B_\nabla$ is increased, hence the droplet stretches more. Importantly, we notice that the potential (\ref{potential_aspect_ratio_2}) is not bounded from below in the high gradient limit $R_0 B_\nabla \gg B_\lambda$. According to equation (\ref{potential_aspect_ratio_2}), there exists a threshold beyond which the aspect ratio diverges to infinity, which is of course non physical. In fact, the limitation to the gradient and second derivative breaks down when the height of the droplet is very high, since the Taylor expansion of the magnetic field is done in terms of $z$. Higher order terms in $h^n,n>3$ should be taken into account in order to counter the squared gradient contribution and suppress the divergence. Note however that gradients created by usual magnets are much too weak to encounter such issue.

Table \ref{tab:Table_1} provides orders of magnitudes for the four dimensionless numbers, from the data provided by the work of \citet{Timonen2013}. We notice that $B_\nabla$ can indeed be neglected, as well as $\tilde{\chi}$ if the field is not very small. Note however that both terms will be taken into account as additional corrections in the numerical resolution of section \ref{sect:numerical_resolution}.

Neglecting $B_\nabla$, the minimization of the potential with respect to $K$ gives:
\begin{equation}
0 = - R_0^2 B_\lambda K^{-5/3}
+ f'(K)
+ \frac{R_0}{3} N'(K) \left( \frac{\chi B_{m,1}}{(1 + N(K) \chi)^2}
+ \frac{\tilde{\chi} B_{m,2}}{(1 + N(K) \tilde{\chi})^2} \right)
\label{potentiel_minimization_K}
\end{equation}

For $\tilde{\chi} \ll \chi$, this can be simplified to\footnote{This is also true in the linear regime $\tilde{\chi} = \chi$ but then the product $N(K) \tilde{\chi}$ can not be neglected anymore.}:
\begin{equation}
R_0 B_m = \frac{(1 + N(K) \chi)^2}{\chi N'(K)} \left( R_0^2 B_\lambda K^{-5/3} - f'(K) \right)
\label{eq_Bm}
\end{equation}
where 
\begin{align}
B_m &\equiv \frac{1}{3} \frac{\mu_0}{\sigma \chi} M(H_a(0))^2
\end{align}
\begin{figure}
\centerline{
\resizebox{0.55\textwidth}{!}{\includegraphics{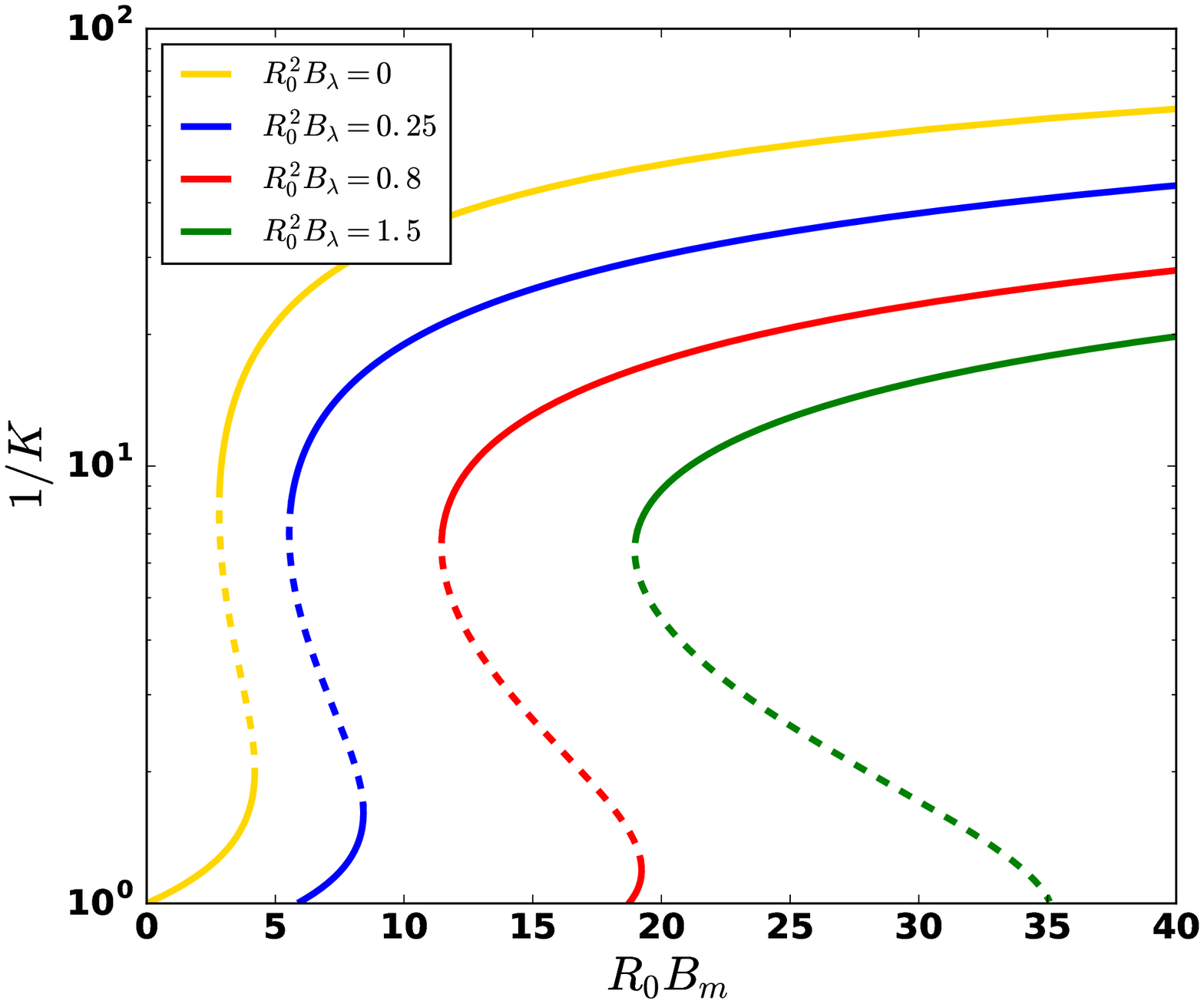}}
\hspace{-0.8cm}
\resizebox{0.55\textwidth}{!}{\includegraphics{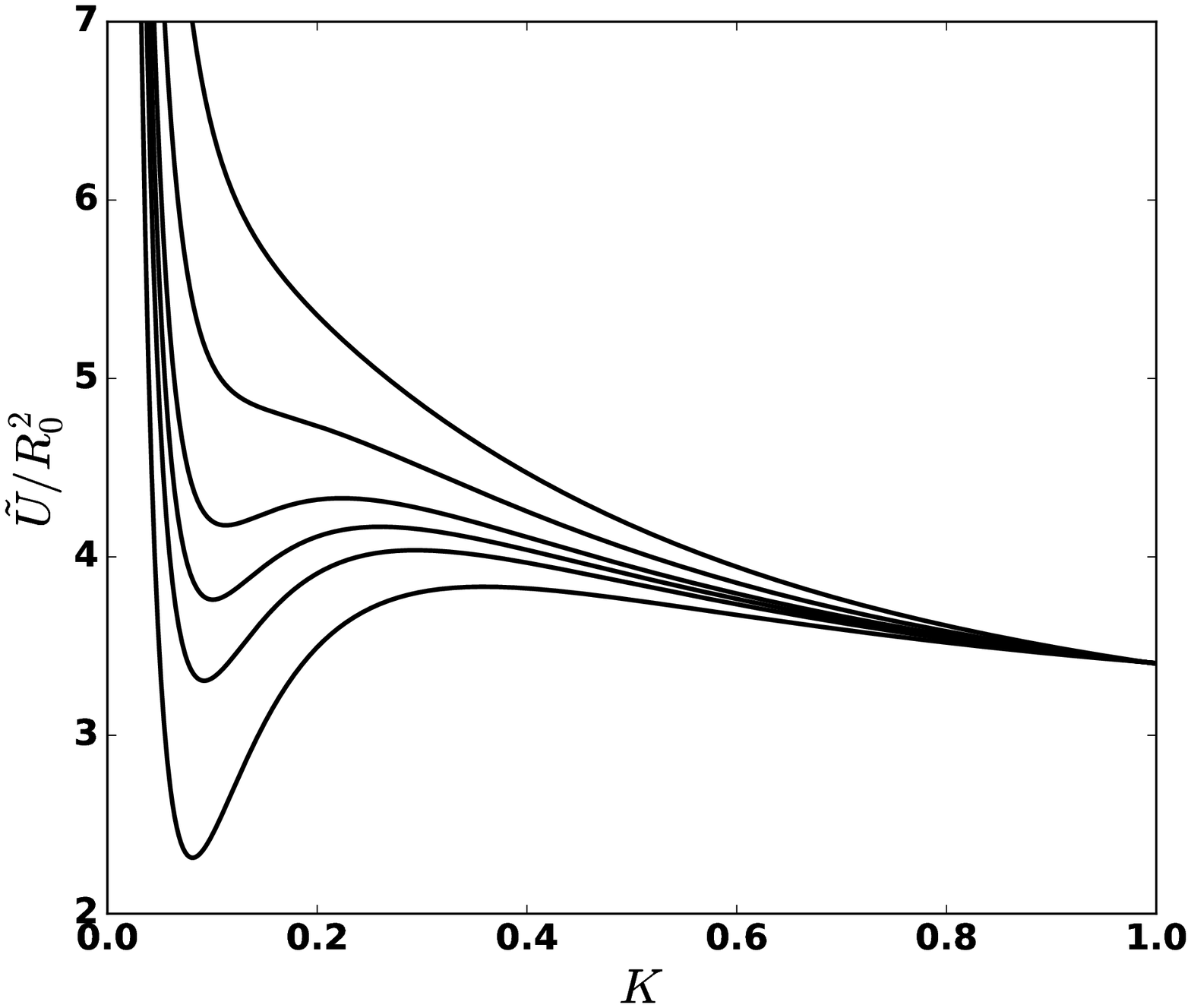}}}
\caption{Left: Modification of the aspect ratio instability for $B_\lambda > 0$, with the apparition of a threshold and the suppression of the stable branch (solid line) at $K \sim 1$ for high values of $R_0^2 B_\lambda$. Here, $\chi = 50$. The yellow curve is the same as the yellow curve of figure \ref{fig:Figure_7} ($\chi = 50$, $R0^2 B_\lambda = 0$). Right: First order phase transition for $\chi = 50$, $R_0^2 B_\lambda = 1.5$ (corresponding to the green curve). From highest to lowest curve: $R_0 B_m = 15, 18, 20, 21, 22, 24$. A shift has been applied to the curves in order to have the same value at $K=1$.}
\label{fig:Figure_8}
\end{figure}

For $B_\lambda = 0$ and $\chi \gg 1$ we retrieve the diagram showing the bistability observed by \citet{Bacri1982}, which is shown in figure \ref{fig:Figure_7}. The aspect ratio increases with the magnetic field strength. For small susceptibilities, this increase is continuous, as in the experimental and numerical study done by \citet{Zhu2011}. For strong susceptibilities, there is a first order phase transition at some point: the aspect ratio jumps from a low value to a higher value. This was observed by \citet{Bacri1982} for a droplet in an immiscible fluid, where the ferrofluid was characterized by an effective susceptibility $\chi = (\mu_2 - \mu_1)/\mu_1$ where $\mu_2$ and $\mu_1$ are respectively the ferrofluid permeability and the surrounding fluid permeability. Such very strong susceptibilities are not unrealistic in the case of very concentrated ferrofluids as the one used by \citet{Bacri1984}.

For $R_0^2 B_\lambda \neq 0$, equation (\ref{eq_Bm}) provides modified diagrams as plotted in figure \ref{fig:Figure_8}. We observe the appearance of a threshold for the control parameter $R_0 B_m$: for a given $R_0^2 B_\lambda \neq 0$, the stable branch does not start at $R_0 B_m = 0$, hence nothing happens ($K$ is equal to 1) until the magnetic field strength reaches the threshold. Beyond this threshold, there is either the same type of first order phase transition as before, for small values of $R_0^2 B_\lambda \neq 0$, or a new transition. In the first case, the first part of the branch is stable, hence the aspect ratio increases continuously until it reaches the unstable branch and jumps to a much higher value. In the second case, once the threshold is reached, the system enters right away into an unstable state, and thence jumps immediately to a much higher aspect ratio. In both cases, before the critical threshold, the second stable equilibrium is separated from the initial one by a barrier of potential, as shown by the right plot of figure \ref{fig:Figure_8}. This barrier decreases as $B_m$ increases until it vanishes when $B_m$ reaches the threshold.

The minimization condition can be rewritten to link the aspect ratio to the parameter $B_\lambda$:
\begin{equation}
R_0^2 B_\lambda = \left(R_0 B_m \frac{\chi N'(K)}{(1 + N(K) \chi)^2} + f'(K) \right) K^{5/3}
\end{equation}

\begin{figure}
\centerline{\hspace{0.25cm}
\resizebox{0.55\textwidth}{!}{\includegraphics{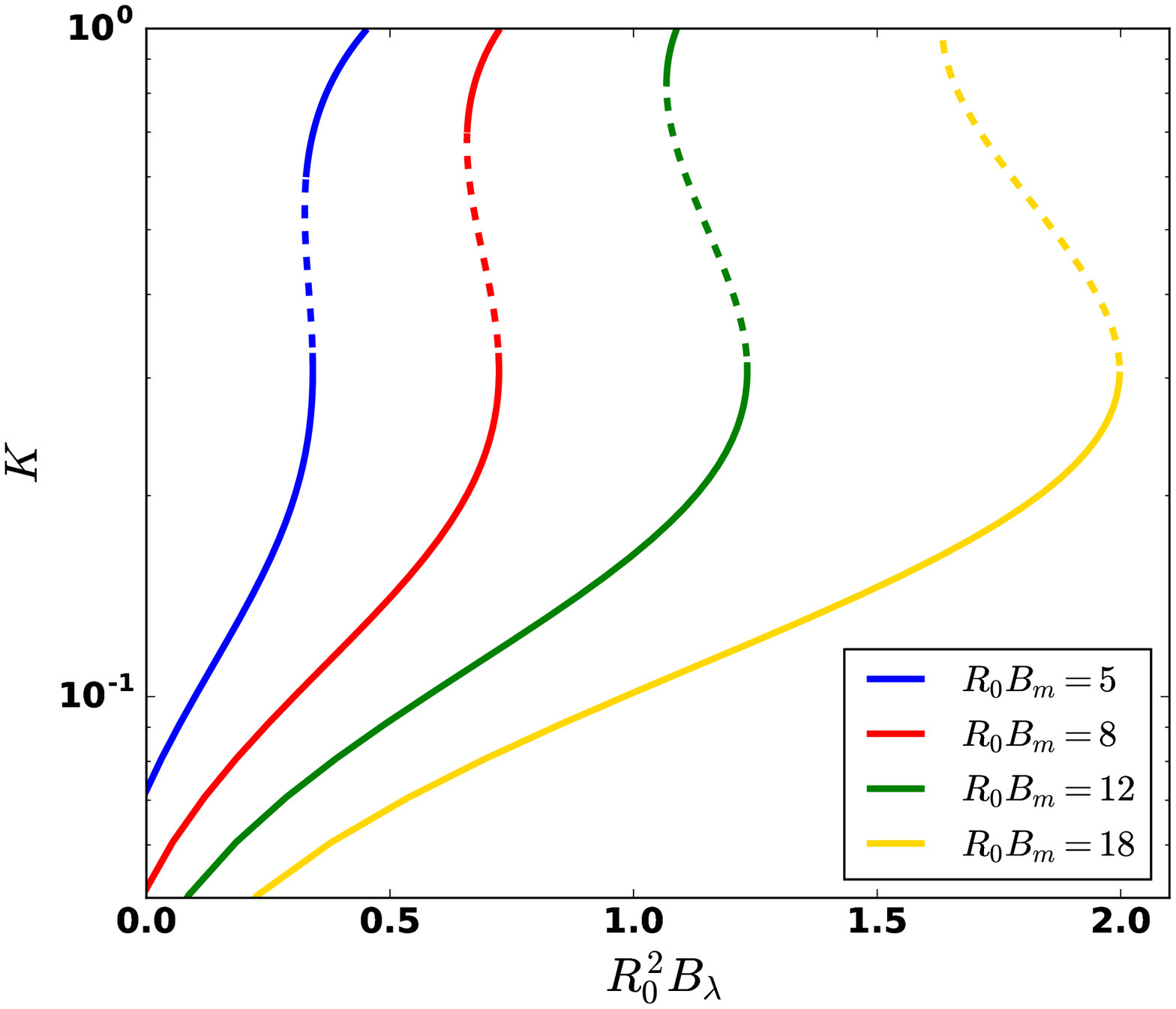}}
\hspace{-0.5cm}
\resizebox{0.55\textwidth}{!}{\includegraphics{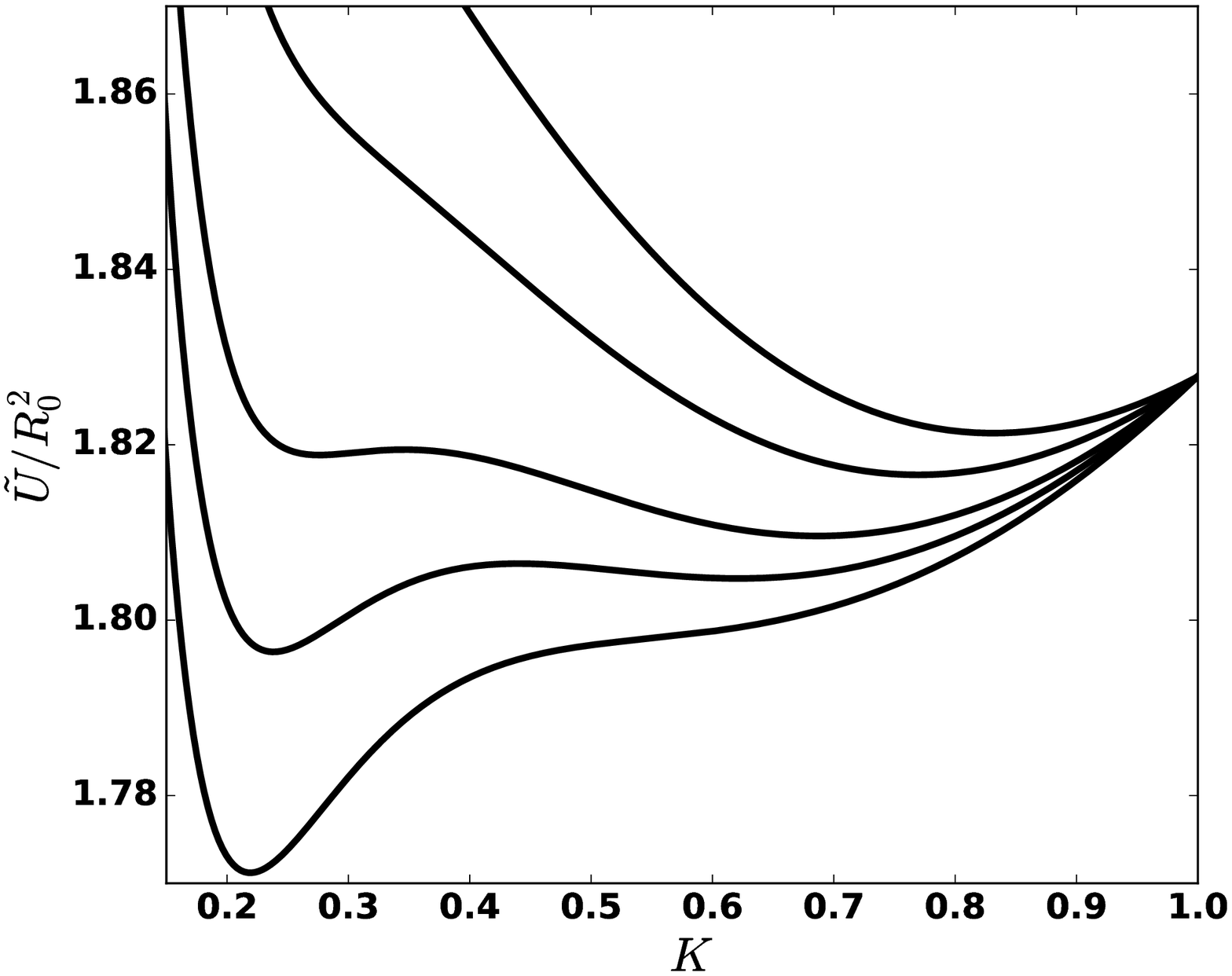}}}
\caption{Left: Instability for $B_\lambda > 0$, with $\chi = 20$. Solid line parts are stable, dotted line parts are unstable. Right: First order phase transition for $\chi = 20$, $R_0 B_m = 5$ (corresponding to the blue curve). From lowest to highest minimum: $R_0^2 B_\lambda = 0.32, 0.33, 0.34, 0.36, 0.38$. A shift has been applied to the curves in order to have the same value at $K=1$.}
\label{fig:Figure_9}
\end{figure}

The left plot of figure \ref{fig:Figure_9} shows the corresponding diagram, i.e. the evolution of the inverse aspect ratio $K$ as function of the control parameter $R_0^2 B_\lambda$. The droplet flattens as $R_0^2 B_\lambda$ increases, and may reach an unstable branch: we therefore identify a third possibility of first order phase transition, now with $R_0^2 B_\lambda$ as control parameter. Although the droplet starts to flatten continuously, it suddenly collapses to a shape closer to that of a sphere ($K=1$) at some point, and then evolves continuously towards a spherical shape as shown by figure \ref{fig:Figure_9}. For high enough values of $R_0 B_m$ the droplet directly collapses to $K \geq 1$ once the instability is reached. Since the potential (\ref{potential_aspect_ratio_2}) was derived assuming a prolate ellipsoidal shape, it is not possible for now to investigate the behaviour for $K > 1$. It will be the purpose of the next section, where we will see that the transition in fact occurs from a prolate ($K<1$) to an oblate ($K>1$) ellipsoid, with nothing special about the spherical shape ($K=1$).

This third phase transition corresponds to the one described in the previous section in the case of a thin layer of ferrofluid, and shown by figure \ref{fig:Figure_5}. It is interesting to note the similarity of those two first order phase transitions despite the discrepancies in the considered system and the employed formalism.

\begin{figure}
\centerline{
 \hspace{0.55cm}
  \resizebox{0.6\textwidth}{!}{\includegraphics{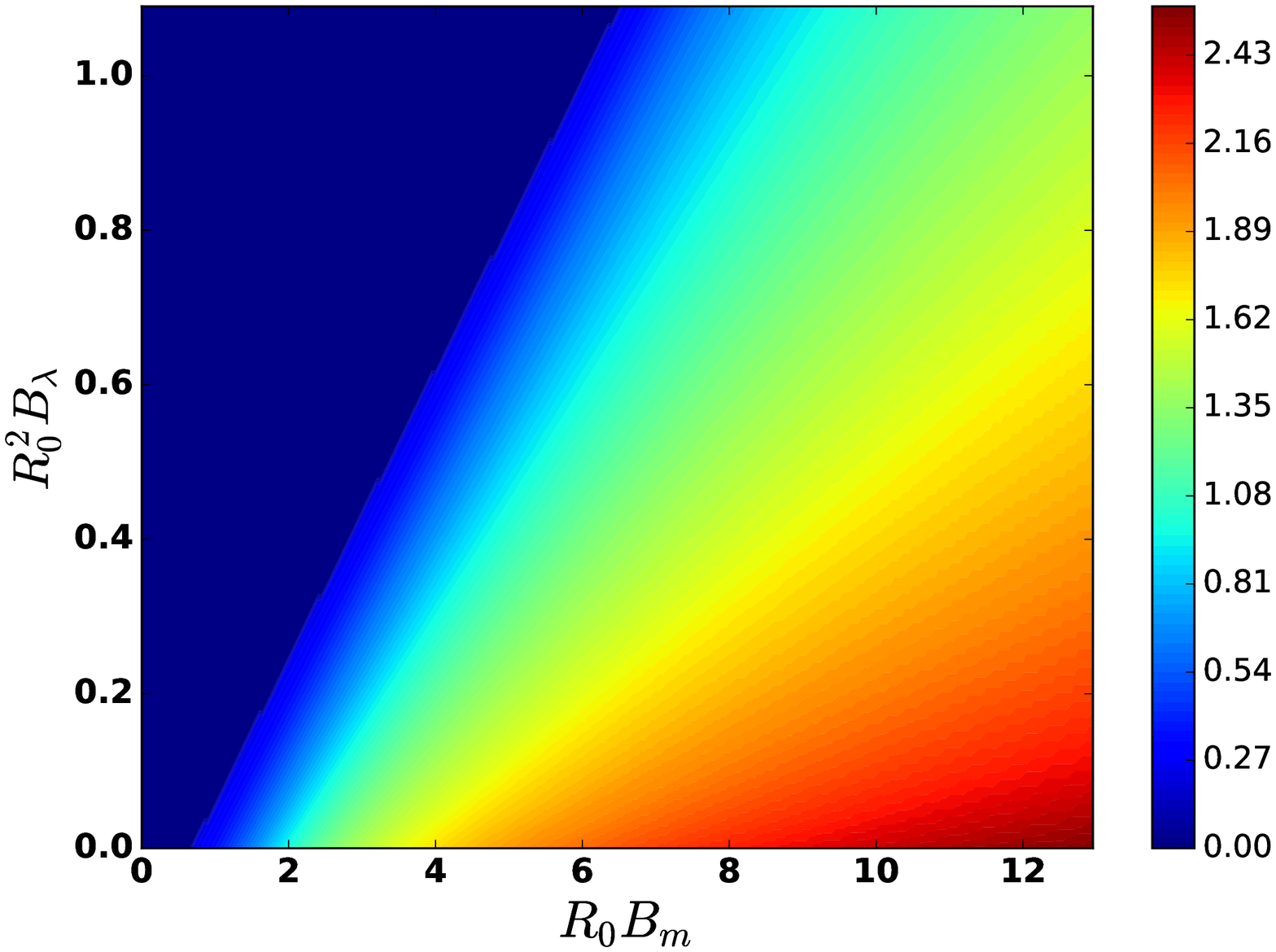}}
 \hspace{-1.2cm} 
  \resizebox{0.6\textwidth}{!}{\includegraphics{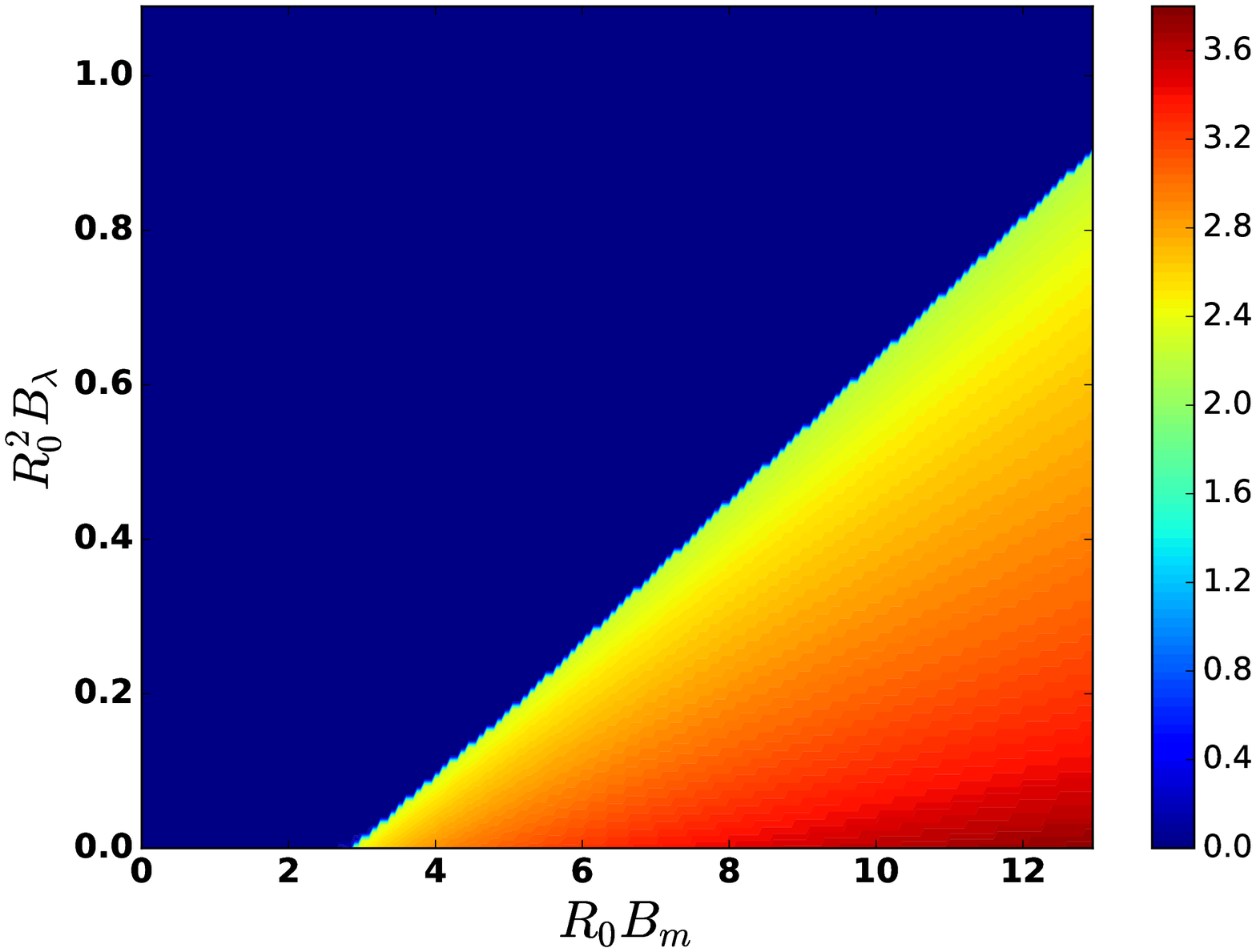}}}
\caption{Phase diagram for $\chi = 5$ (left) with second order phase transition, and for $\chi = 50$ (right) with first order phase transition. The colors correspond to the value of $\log 1/K$ computed at each couple of parameters $(R_0 B_m,R_0^2 B_\lambda)$ by numerically inverting equation (\ref{eq_Bm}). The deep blue area corresponds more generally to an oblate ellipsoid ($K \geq 1$).}
\label{fig:Figure_10}
\end{figure}

Figure \ref{fig:Figure_10} shows the phase diagram $(R_0 B_m, R_0^2 B_\lambda)$, i.e. the value of the aspect ratio $1/K$ corresponding to the deepest minimum of the potential for each couple of parameters $(R_0 B_m, R_0^2 B_\lambda)$. A single transition line is evidenced. For small susceptibilities, the transition is second order (e.g. blue and red curves of figure \ref{fig:Figure_7}), i.e. the aspect ratio starts increasing or decreasing continuously as the transition line is crossed. For high susceptibilities, the transition is first order and corresponds to the previous diagrams $(1/K,R_0 B_m)$ and $(K,R_0^2 B_\lambda)$ (figures \ref{fig:Figure_8} and \ref{fig:Figure_9}). We see here that the previously evidenced transitions correspond in fact to the same transition, namely the crossing of the line in the phase diagram.
It can be either crossed from low aspect ratio to high aspect ratio, increasing $B_m$, or from high aspect ratio to low aspect ratio, increasing $B_\lambda$:
the gradient contribution enhances the effect of gravity and flattens the droplet, while the demagnetization contribution tends to increase the interface area.

Note that, in general, the behaviour of the droplet as function of an approaching magnet will be a combination of the previous conclusions, following a trajectory $(R_0 B_m, R_0^2 B_\lambda)$ in the phase diagram, which may cross the phase transition line several times. Numerical examples of such realistic processes will be shown in section \ref{sect:numerical_resolution}.
It would be very interesting to design future experiments to investigate the predicted transition. One could for instance use a combination of two fields: one uniform and the other non-uniform. A very precise control of the increase of the magnetic field at low field would probably be required, as well as a very good hyperhydrophobic surface (or, better, an immiscible fluid).
We can predict an experimental observation in 4 steps as the magnet approaches. First, the droplet does not react because the field is below the initial threshold. Second, the droplet stretches to an aspect ratio which can be quite high depending on the field configuration. Third, the droplet shape suddenly collapses to an oblate ellipsoid. Fourth, the aspect ratio slowly decreases until fissions occur. It should be possible to observe more complicated scenarii with several successive transitions, as we will see in section \ref{sect:numerical_resolution} through a numerical resolution.

To conclude this section, we show a quantitative comparison of our predicted aspect ratios and the data obtained by \citet{Timonen2013}. Figure S4 of \citet{Timonen2013} provides pictures of droplets submitted to a given field configuration. From these pictures we estimated the aspect ratio of the droplets and compared it to our theoretical prediction (taking into account the gradient squared term and the second derivative of the magnetic field in the term $B_\nabla$). For the few cases where the number of created droplets is bigger than 1, we use the generalized formula which will be derived in the next section. The results, shown in table \ref{tab:Table_1}, prove a good agreement, keeping in mind that the experimental aspect ratios are estimates. It is interesting to notice the discrepancy for high value of $R_0^2 B_\lambda$: the theory predicts an oblate ellipsoid (the phase transition has happened), although the aspect ratio remains bigger than 1 in the observations. This may be due to neglected contributions, like the inhomogeneity of the demagnetization which becomes non-negligible for high gradients. A thorough experimental study of the previously evidenced instabilities remains necessary to properly test our theoretical predictions, although those first results are encouraging.

\begin{table}
\begin{center}
\def~{\hphantom{0}}
\begin{tabular}{cc|ccccc|ccc}
$H_a$ [kA/m] & $\left. \frac{\partial H_a}{\partial z} \right|_0$ [kA/m$^{2}$] & $\tilde{\chi}$ & $R_0 B_{m,1}$ & $R_0 B_{m,2}$ & $R_0^2 B_\lambda$ & $10^3 R_0^3 B_\nabla$ & $n$ & $1/K$ exp & $1/K$ theory \\ 
& & & & & & & & & \\
7.16 & 15.9 & 2.7 & 2.9 & 4.0 & 0.011 & 0.08 & 1 & 1.8 & 1.4 \\
7.16 & 398  & 2.7 & 2.9 & 4.0 & 0.27  & 50   & 1 & 1   & 1.1 \\
26.3 & 47.7 & 0.8 & 21  & 54  & 0.077 & 0.21 & 1 & 5 & 5.2 \\
26.3 & 2228 & 0.8 & 21  & 54  & 3.2   & 454  & 7 & 2.4 & 2 \\
52.5 & 103  & 0.2 & 37  & 177 & 0.2   & 0.30 & 1 & 6   & 6.6 \\
52.5 & 5491 & 0.2 & 37  & 177 & 10.7  & 838  & 1 & 1.8 & $<1$ \\
54.1 & 103  & 0.2 & 38  & 185 & 0.2   & 0.28 & 1 & 6   & 6.7 \\
54.1 & 5730 & 0.2 & 38  & 185 & 11.3  & 860  & 2 & 2   & $<1$ \\
70.8 & 135  & 0.1 & 43  & 277 & 0.28  & 0.29 & 1 & 7.1 & 6.9 \\
70.8 & 7958 & 0.1 & 43  & 277 & 16.4  & 1000 & 4 & 1.6 & $<1$ \\
101  & 199  & 0.07 & 50  & 452 & 0.43 & 0.3  & 1 & 5.9 & 6.2 \\
101  & 12335& 0.07 & 50  & 452 & 26.5 & 1200 & 7 & 1.7 & $<1$ \\
\end{tabular}
\caption{Experimental data are estimated from the work of \citet{Timonen2013}. Theoretical values are taken from solutions of equation (\ref{potentiel_minimization_K}) in which the term $B_{\nabla}$ was not neglected  (numerical resolution without any assumption, also taking into account the varying demagnetization in $\tilde{F_M}$ and the variation of $\tilde{\chi}$ although they have no significant impact). The susceptibility of the ferrofluid is $\chi = 3.3$. The initial volume is $V_0 = 20$ $\umu$L and the corresponding spherical radius is $R_0 = 1.7$ mm.}
\label{tab:Table_1}
\end{center}
\end{table}

\subsection{Fate of the spherical droplet}
\label{sect:fate}
We have seen that when the magnetic gradient increases faster than the uniform contribution, the aspect ratio eventually reaches 1. The remaining question is the following: what happens next? One may guess that the gradient will continue to flatten the droplet, transiting from a prolate to an oblate ellipsoid ($K>1$). It is not possible to properly answer this question using the potential (\ref{potential_aspect_ratio_2}) because it has been derived assuming a prolate ellipsoidal shape. However, it can be easily adapted using the analytic continuation of the previous geometrical formula. In particular, the eccentricity, surface, and demagnetization factor are given by \citep[for the demagnetization, see][]{Beleggia2006}:
\begin{gather}
e = \sqrt{1-1/K^2} \\
S_{oblate} = 2 \upi R_0^2 K^{2/3} \left(1 + \frac{1}{e K^2}\tanh^{-1} e \right) \\
N = \frac{1}{e^2} \left(1- \frac{1}{e K} \arcsin e \right)
\end{gather}

Discarding again the $B_\nabla$ term and assuming $\tilde{\chi} \ll \chi$, the potential has the same shape as before:
\begin{equation}
\frac{\tilde{U}(K)}{R_0^2} = \frac{3}{2} R_0^2 B_\lambda K^{-2/3}
+ f_{oblate}(K)
- \frac{R_0 B_{m}}{1 + N(K) \chi}
\label{oblate_droplet_pot}
\end{equation}
where
\begin{equation}
f_{oblate}(K) \equiv K^{-1/3} \left(K + \frac{1}{e K}\tanh^{-1} e \right)
\end{equation}

\begin{figure}
  \centerline{\includegraphics[width=\linewidth]{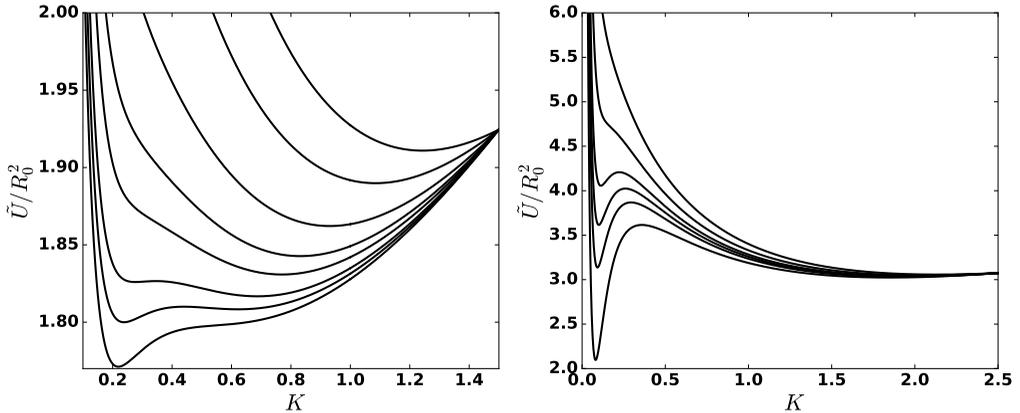}}
\caption{Left: Evolution of the potential for $\chi=20$, $R_0 B_m = 5$ and, from bottom curve to top curve, $R_0^2 B_\lambda=0.32,0.33,0.34,0.36,0.38,0.42,0.5,0.6$ (continuation of the plot of figure \ref{fig:Figure_9}). This shows a prolate-prolate first order phase transition followed by a smooth evolution from a prolate to an oblate shape of the droplet. 
Right: Evolution of the potential for $\chi=50$, $R_0^2 B_\lambda = 1.5$ and, from top curve to bottom curve, $R_0 B_m=15,18,20,21,22,24$ (continuation of the plot of figure \ref{fig:Figure_8}). This shows a prolate-oblate first order phase transition which jumps over the spherical shape. To improve the readability, a shift has been applied to both plots in order to have the same value at $K=1.5$ (left) and $K=2.5$ (right).}
\label{fig:Figure_10_5}
\end{figure}

Figure \ref{fig:Figure_10_5} shows the continuation of the plots of figures \ref{fig:Figure_8} and \ref{fig:Figure_9}. It reveals that the value $K=1$ has in fact nothing special and the aspect ratio keeps evolving after the transition: the gradient indeed keeps flattening the droplet. Moreover, in general, the previously evidenced first order phase transition happens between a prolate to an oblate ellipsoid, hence jumps over the spherical shape.

Basically, we therefore identified a transition from a prolate shaped to an oblate shaped droplet, which should be observable. Note however that we have not yet taken into account the possibility of the droplet's fission, which can occur before the prolate-oblate transition. The study of the fission phenomenon is the purpose of the next section.

\subsection{Fission instability}
\label{sect:fission}
In the previous section we have written the potential of a single ferrofluid droplet (equation (\ref{potential_aspect_ratio_2})). It can be generalized to an ensemble of $n$ droplets of volume $V/n$ and spherical radius $n^{-1/3} R_0$:
\begin{multline}
\qquad \qquad \frac{\tilde{U}(K,n)}{R_0^2} = \frac{8 \upi^2 n^{-1/3} R_0^2}{3} \frac{K^{-2/3}}{\lambda^2}
+ n^{1/3} f(K)
- \frac{2 n^{-2/3} R_0^3}{5} \frac{B_\nabla K^{-4/3}}{1 + N(K) \tilde{\chi}} \\
- \frac{R_0}{3} \left( \frac{B_{m,1}}{1 + N(K) \chi} 
+ \frac{B_{m,2}}{1 + N(K) \tilde{\chi}} \right) \qquad \qquad \qquad \qquad 
\label{potential_n_droplets}
\end{multline}

Starting from a single droplet, it is energetically favorable to fission the initial droplet if the following ``rupture condition'' holds:
\begin{equation}
U(K_1,n=1) - U(K_2,n=2) > 2 \upi \epsilon R_0^2
\label{rupture_condition_0}
\end{equation}
where $\epsilon$ is a phenomenological energy per unit surface taking into account the irreversibility of the phenomenon. It can be put to zero in first approximation.

Equations (\ref{potential_n_droplets}) and (\ref{rupture_condition_0}) gather two mechanisms capable of splitting a droplet of magnetic fluid. When the field is uniform, the last term of equation (\ref{potential_n_droplets}) can trigger the instability through the variation of the demagnetization factor, providing that the initial volume is sufficiently high. This topological instability in uniform fields has been evidenced in the experiments by \citet{Barkov1980} and theorized by \citet{Berkovsky1985} using an energetic approach similar to ours. An important conclusion of these works is that there exists a minimum initial volume allowing the fission, typically of order 0.1 mL for usual magnetic intensities and fluid properties \citep{Berkovsky1987,Berkovsky1993,Blums1997}. This is why the splitting effect was not observed in the experiments of \citet{Timonen2013}, who stated that it is generically not observable with uniform magnetic fields, which is only true for small droplets. In the case of droplets of typical volume 10 $\umu$L, the variation of the last term of equation (\ref{potential_n_droplets}) is indeed too small to trigger the fission of the initial droplet, and the key mechanism becomes the magnetic force created by the non-uniformity of the field, i.e. the first term of equation (\ref{potential_n_droplets}).

In the following, we therefore assume for simplicity that $K_1 = K_2$, i.e. the aspect ratio evolves continuously while the fission occurs. We computed that the discontinuity, if it exists, is less than 0.1, hence does not introduce any significant correction. This implies that the last term of equation (\ref{potential_n_droplets})
is negligible and the fission phenomenon is indeed observable only when the applied field is non-uniform, as experimentally concluded in \citet{Timonen2013}.

Equation (\ref{rupture_condition_0}) reduces to the following condition:
\begin{align}
\frac{\lambda^2}{R_0^2} < \left( \frac{4 \upi \mathcal{C}(K)}{3} \right)^{2/3}
\label{rupture_condition}
\end{align}
where
\begin{equation}
\mathcal{C}(K) =  \frac{4\upi^2 \sqrt{2/3}}{K} (1-2^{-1/3})^{3/2}
\left( \frac{\epsilon}{\sigma} + (2^{1/3}-1) f(K) + (1-2^{-2/3})\frac{2 R_0^3}{5} \frac{B_\nabla K^{-4/3}}{1 + N(K) \tilde{\chi}} \right)^{-3/2}
\end{equation}

For $B_\nabla =0$, $\epsilon = 0$, $K=1$, $\mathcal{C}(K) \simeq 8$ and equation (\ref{rupture_condition}) is the fission criterion intuited by \citet{Timonen2013}: the droplet fissions if the characteristic wavelength is smaller than the initial diameter.

Starting with a distribution of $n$ droplets of spherical radius $n^{-1/3} R_0$, the rupture condition becomes:
\begin{equation}
\frac{\lambda^2}{R_0^2} < n^{-2/3} \left( \frac{4 \upi \mathcal{C}(K)}{3} \right)^{2/3}
\end{equation}

Theoretically, if this condition holds, we should obtain a distribution of $2n$ droplets\footnote{In practice there is always one droplet bigger than the others, hence with a bigger $R_0$, hence which splits first.}. Hence the number of droplets for a given $\lambda$ is:
\begin{align}
n  = \left\lfloor \frac{\mathcal{C}(K_c)}{2} \frac{V_0}{\lambda^3} \right\rfloor
\label{number_droplets}
\end{align}
where $V_0$ is the initial volume of fluid, $K_c$ is the inverse aspect ratio which minimizes the potential at the fission.

The theoretical number of droplets predicted by equation (\ref{number_droplets}) does not depend on the uniform contributions $B_{m,1}$ and $B_{m,2}$: the creation of droplets is essentially driven by the competition between surface tension, which tends to keep a small number of droplets, and the gradient, which tends to create as much as small droplets as possible. The squared gradient tends to elongate the droplets: it introduces a correction which slows down the droplets fission.

\begin{figure}
  \centerline{\includegraphics[width=0.75\linewidth]{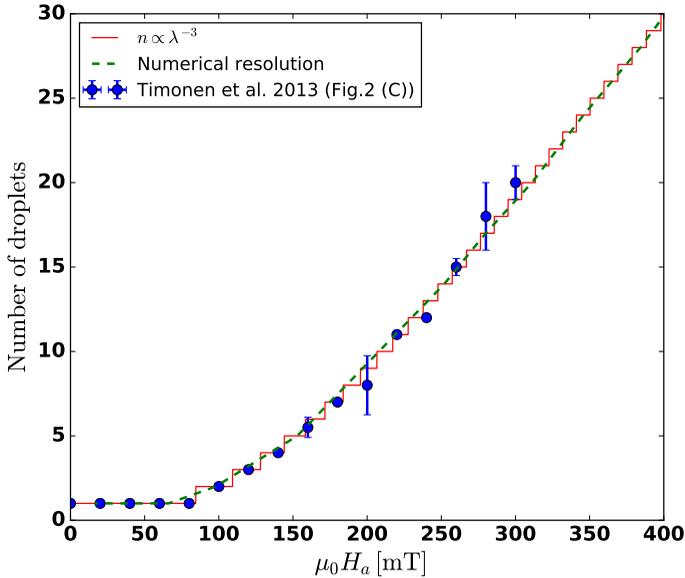}}
\caption{Comparison between our theoretical prediction for the evolution of the number of created droplets and the experimental data (blue points) taken from \citet{Timonen2013}. The red line shows the agreement obtained with the proportionality law $n \propto \lambda^{-3}$ parametrized using the corresponding parameters (magnet M3) given in the Supplementary Materials of \citet{Timonen2013}, and with the slope as fitting parameter. The green dashed line show the theoretical prediction obtained by taking into account $B_\nabla$ in equation (\ref{number_droplets}) and a supplementary global factor of 0.75.}
\label{fig:Figure_11}
\end{figure}

The relation $n \propto V_0 \lambda^{-3}$ is in agreement with the experimental results of \citet{Timonen2013}. They indeed identified this proportionality up to a large number of created droplets (see figure S2D of the supplementary materials of \citet{Timonen2013}). This suggests first that the coefficient $\mathcal{C}(K_c)$ is only weakly dependent of $n$, i.e. $K_c$ is an almost scale-invariant criterion, and second that the contribution from $B_\nabla$ is small. In first approximation, the slope of the law $n \propto \lambda^{-3}$ is indeed constant and moreover universal in the sense that it does not depend on the field configuration or fluid properties.
This explains why the slopes are roughly the same in the vicinity of the origin in figure S2D of \citet{Timonen2013}. We can for instance adjust the experimental data provided by figure 2C of \citet{Timonen2013}, with $\lambda$ given by equation (\ref{equation_lambda}) and a free coefficient as fitting parameter. Figure \ref{fig:Figure_11} shows the comparison between these data (blue points) and the proportionality law (red curve). The global factor is 0.38. This factor can be theoretically obtained taking $\epsilon \simeq 0.5 \sigma$ in equation (\ref{rupture_condition_0}), which means that the energetic cost of the irreversibility should be about five times less than the total surface tension energy of the initial droplet, which is big. We will however show in the next section that the squared gradient contribution, taken into account in the green line fit, brings the global factor closer to 1.

We observe in \citet{Timonen2013}, that the slope of equation (\ref{number_droplets}) is in fact not universal: it weakly depends on the magnet. This is due to the squared gradient contribution which is not negligible anymore and slows down the fission process. One can check that the hierarchy of the slopes in figure S2D of \citet{Timonen2013} is indeed in agreement with the hierarchy of the gradients. For a larger number of droplets, the proportionality is not even verified anymore: non-linear deviations are observed. The numerical resolution of the next section will show that those are not due to the squared gradient contribution. We suspect that they are due to the radial variation of the field in the horizontal plane since for a large number of droplets the droplets are less confined, hence not aligned with the center of the magnet anymore.

\subsection{Numerical resolution}
\label{sect:numerical_resolution}
\begin{figure}
  \centerline{
    \includegraphics[width=0.55\linewidth]{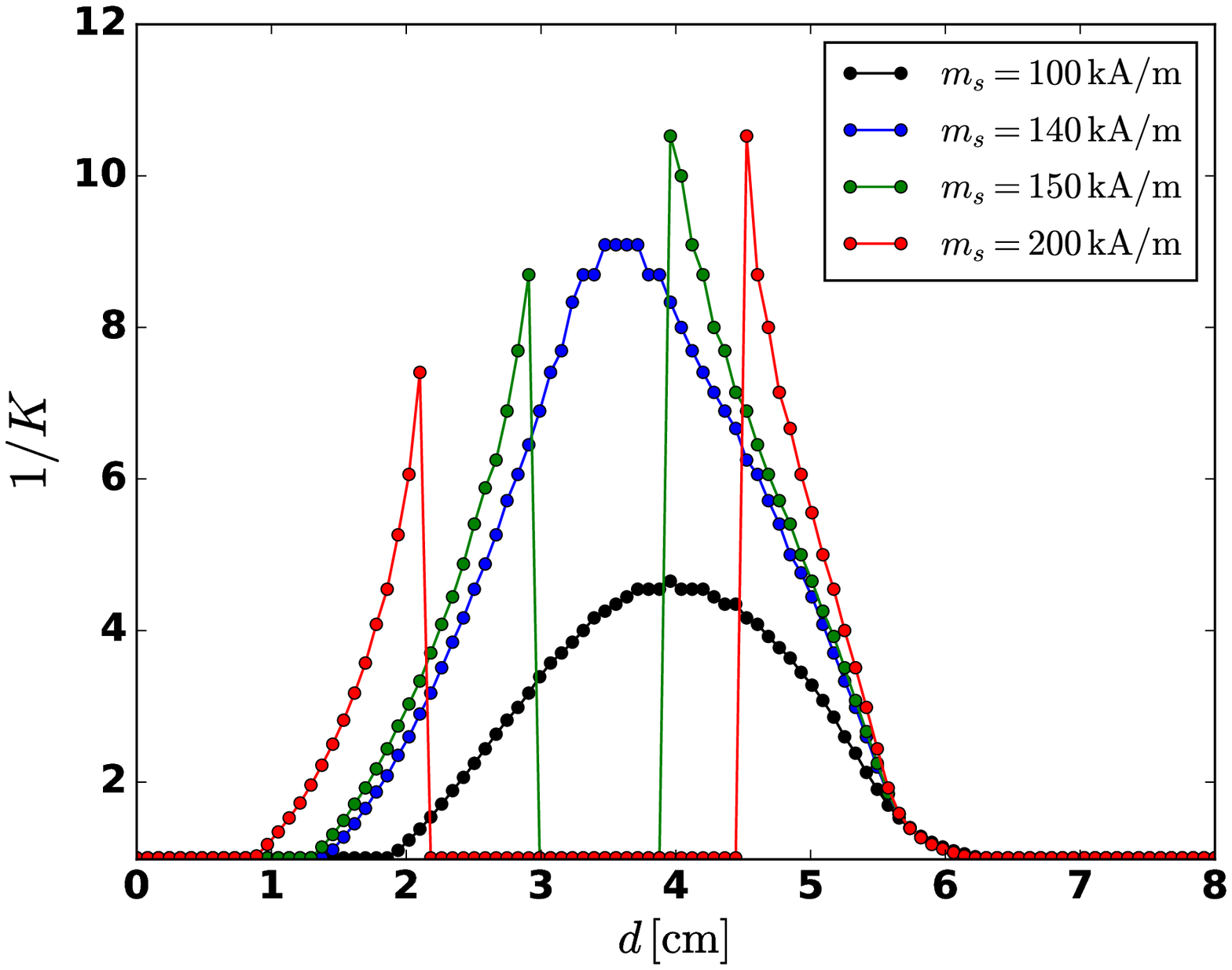}}
  \centerline{
    \resizebox{0.55\linewidth}{!}{\includegraphics{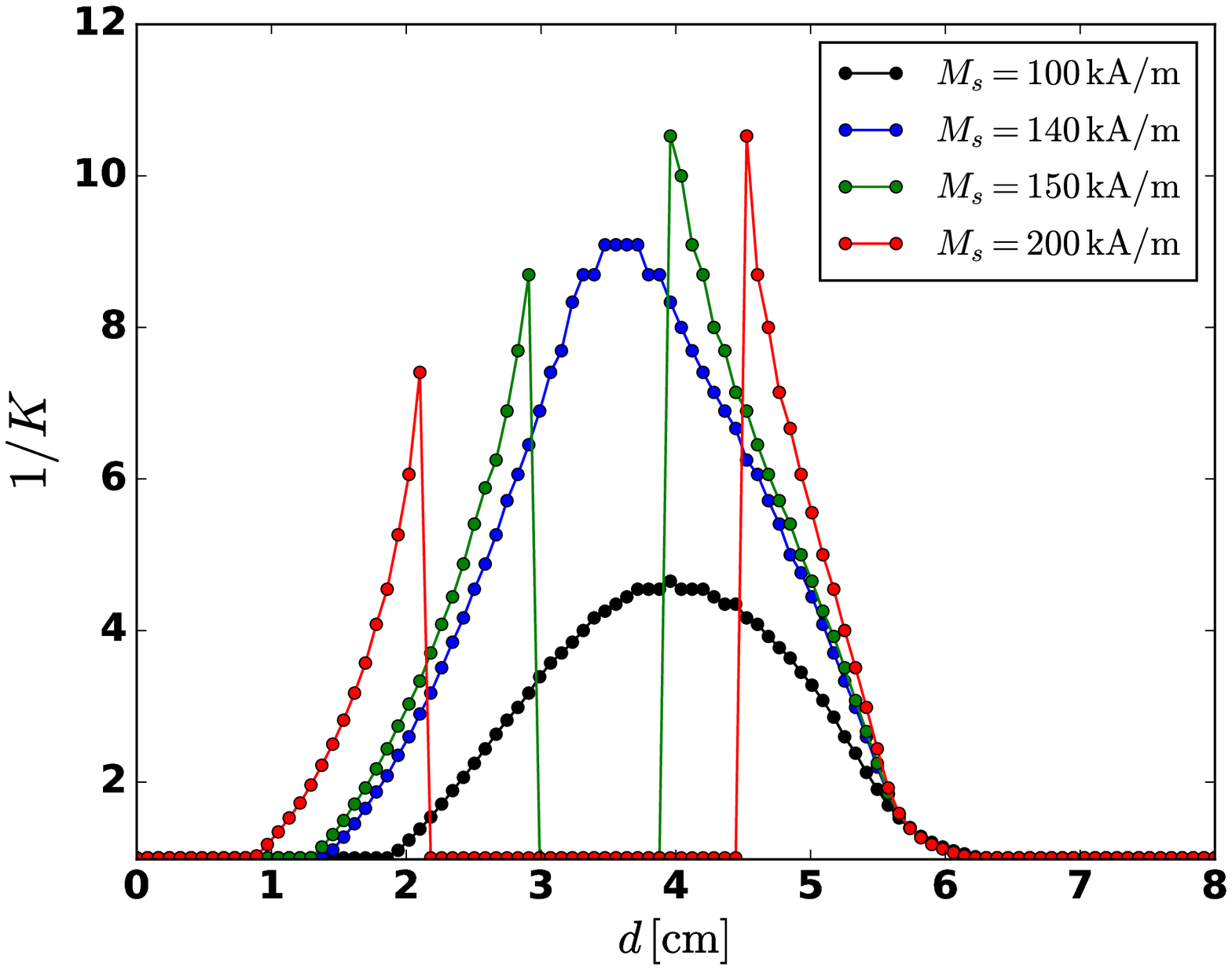}}
    \hspace{-0.6cm}
    \resizebox{0.55\linewidth}{!}{\includegraphics{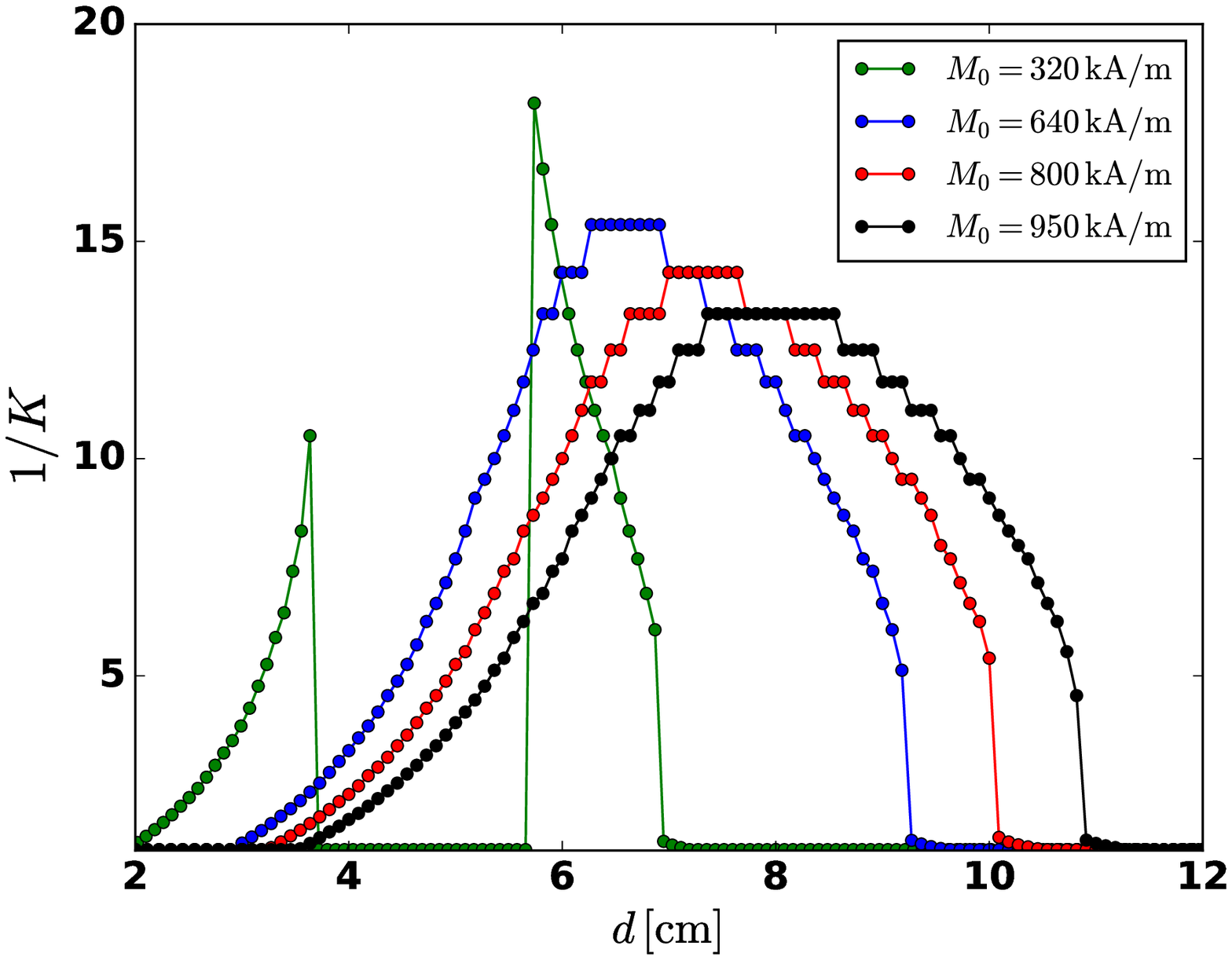}}}
\caption{Numerical resolution for the evolution of the aspect ratio, obtained by solving equation (\ref{potentiel_minimization_K}), taking into account both $B_{m,1}$ and $B_{m,2}$ and also taking into account the squared gradient and the second magnetic field derivative in $B_\nabla$. Top: $\chi=10$, $M_0=320$ kA/m. Left: $M_s = 200$ kA/m, $M_0=320$ kA/m. Right: $\chi = 40$, $M_s = 200$ kA/m. Between the top and left plots the red curve is the same. Between the left and right plots the green curve is the same.}
\label{fig:Figure_12}
\end{figure}

In a first part, we solve equation (\ref{potential_aspect_ratio_2}) numerically in order to obtain the evolution of the aspect ratio $1/K$ as function of the field configuration, which evolves when decreasing the distance between the fluid and the magnet. The squared gradient and second magnetic field derivative are taken into account in $B_\nabla$. $B_{m,1}$ and $B_{m,2}$ are both taken into account without further assumption. In the first plot of figure \ref{fig:Figure_12}, we use as input a ferrofluid of susceptibility $\chi=10$, surface tension $\sigma = 50$ mN/m, density $\rho = 2000$ kg/m$^3$ and various saturated magnetizations.  The magnetic field is computed using equation (\ref{magnetic_field_Camacho}) with a radius $R=2$ cm, a height $h=4$ cm, and an intrinsic magnetization $M_0=320$ kA/m (e.g. a ferrite magnet). The droplet has a volume of 10 $\umu$L. As the distance $d$ to the magnet is decreased, for $M_s\lesssim 140$ kA/m, the aspect ratio is observed to evolve continuously, first increasing as the uniform contribution increases and then decreasing once the gradient contribution becomes predominant. But for higher values of the saturated magnetization we see the apparition of a succession of various phase transitions: while the magnet is approached (from right to left on the plot), $1/K$ first becomes bigger than one (prolate ellipsoid) in a second order phase transition and then continuously increases, until $R_0^2 B_{\lambda}$ becomes high enough to have a first order phase transition evidenced in figure~\ref{fig:Figure_9}, i.e. $1/K$ collapses to a value smaller than one (oblate ellipsoid). This transition is caused by the magnetization becoming saturated, hence drastically reducing the increase of the uniform contribution. Then when the contributions of $R_0 B_{m,1}$ and $R_0 B_{m,2}$ become high enough again, a second first order phase transition occurs, of the kind evidenced in figure~\ref{fig:Figure_8}, i.e. $1/K$ suddenly jumps to a higher value, and then decreases again because of the gradient domination, until becoming again smaller than one.

We did the same kind of simulation but fixing the saturated magnetization at $M_s= 200$ kA/m, and increasing the susceptibility, while taking the other parameters unchanged. The results are shown on the second plot of figure \ref{fig:Figure_12}. We still observe the phase transitions described above, but the first one becomes first order: for high enough values of $\chi$, the aspect ratio almost immediately jumps to a higher value (see figure \ref{fig:Figure_8}).

Finally, after fixing $\chi=40$, we increase the intrinsic magnetization of the magnet in the third plot of figure \ref{fig:Figure_12}. We can see that the two middle phase transitions disappear while the first first order phase transition remains (the other small discontinuities are numerical errors).

These numerical results show that taking into account all contributions does not prevent the occurrence of the previously evidenced phase transitions with realistic ferrofluid and magnet parameters, and should therefore be observable. Furthermore, it shows that in realistic situations, all the instabilities and phase transitions can appear as coupled or superposed. Still, it is necessary to check their existence through experimental studies.
%
\begin{figure}
\resizebox{0.5\textwidth}{!}{%
  \includegraphics{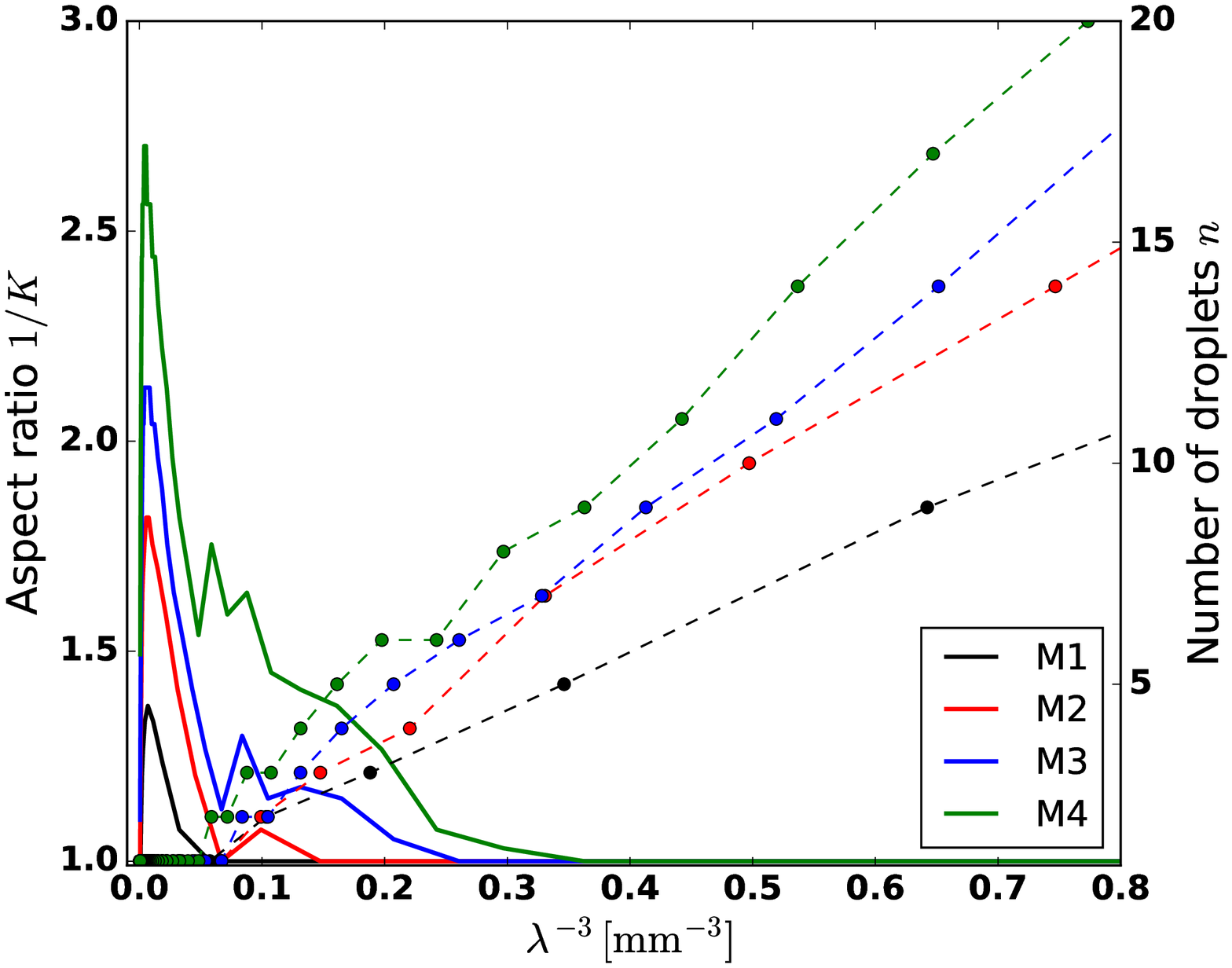}}
\resizebox{0.5\textwidth}{!}{%
  \includegraphics{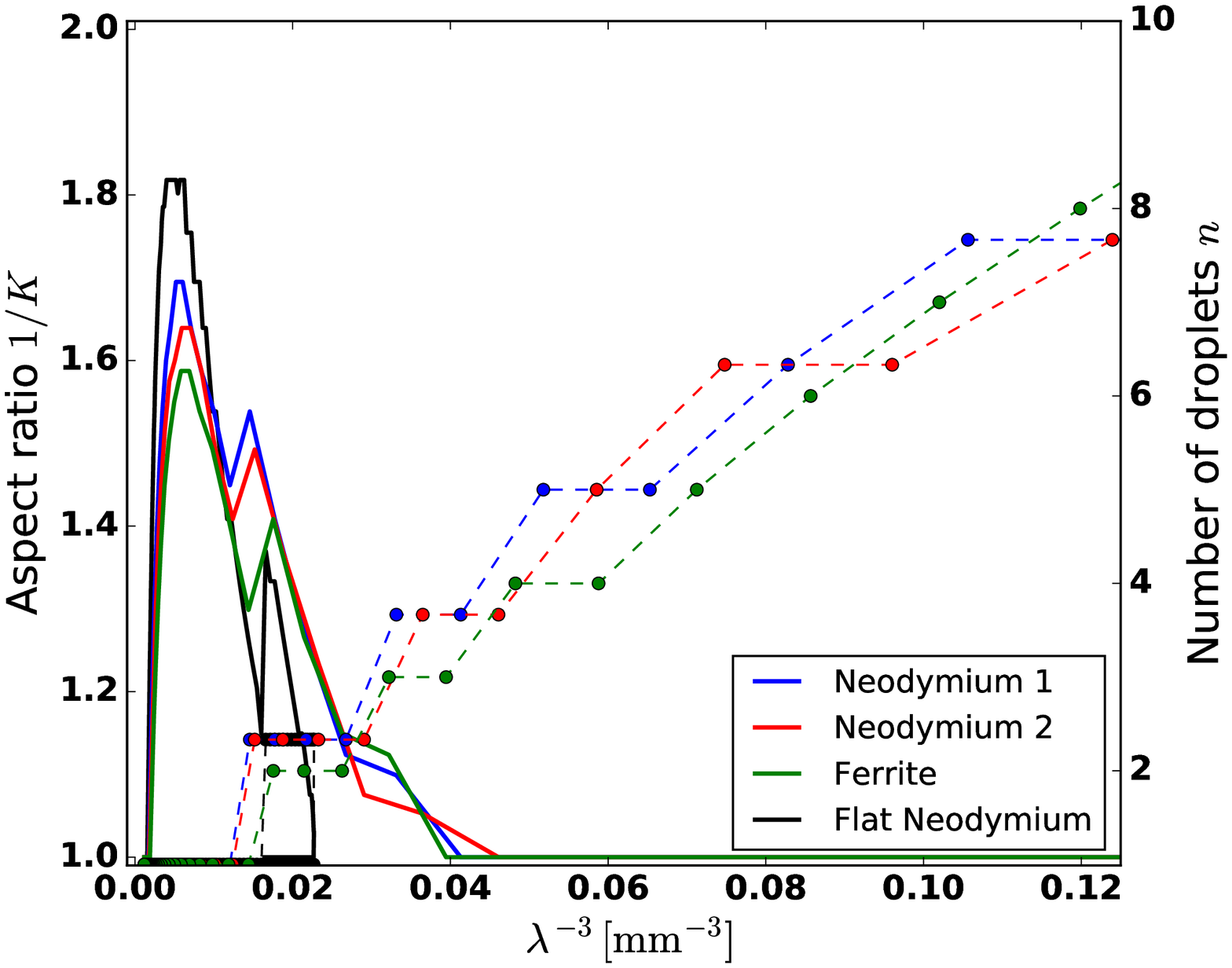}}
\caption{Numerical resolution of the evolution of the aspect ratio (solid lines) and the number of droplets (dotted lines) as functions of $\lambda^{-3}$, for an initial droplet of volume $10~\umu$L. This simulates an approach from $d \sim 8$ cm to $d \sim 0.5$ cm. Left: the input parameters correspond to the experiments from the work of \citet{Timonen2013}, in the cases of their four magnets described in the Supplementary materials. The correction factor is 0.75. Right: the input parameters correspond to our non-diluted ferrofluid and our four magnets described in section \ref{sect:magnets}. The correction factor is 1.35.}
\label{fig:Figure_13}
\end{figure}
%

\medbreak
In a second part, we focus on a numerical study of the fission behaviour. For a given field configuration, the minimization of the potential of equation (\ref{potential_n_droplets}) with respect to $K$ gives the droplets aspect ratio for a fixed number of droplets and equation (\ref{number_droplets}) gives the number of droplets for a fixed aspect ratio.
We simulated the behaviour of a ferrofluid droplet submitted to an increased field by numerically solving this system. As inputs, we used the characteristics of our magnets and magnetization curve (see table \ref{tab:Table_2}, figure \ref{fig:Figure_15}, figure \ref{fig:Figure_14}), as well as the characteristics of the magnets and fluid used by \citet{Timonen2013}.

First we need to calibrate equation (\ref{number_droplets}). In the case of the magnets of Timonen \textit{et al.}, this is done by correcting the numerical solution obtained in the case of their magnet M3, comparing it to the experimental data of Timonen \textit{et al.} and the simple proportionality law plotted in figure \ref{fig:Figure_11}. The numerical resolution corresponds to the green dotted line. The fit is obtained by introducing a factor equal to 0.75 multiplying $\mathcal{C}(K_c)$ in equation (\ref{number_droplets}). This is equivalent to taking $\epsilon \simeq 0.1 \sigma$, hence an irreversibility about 20 times smaller than the surface tension energy, which is realistic and shows that the loss of energy due to the irreversibility may satisfactorily explain the small discrepancy between the experimental and predicted factors.
In the case of our magnets, the factor is determined by comparing the numerical resolution to the experiments which will be performed in section \ref{universal_behaviour}. We need to multiply $\mathcal{C}(K_c)$ by 1.35 (the same factor for our four magnets). We attribute the difference with the factor 0.75 obtained for the magnets of \citet{Timonen2013} to the difference of geometry. Since we are working on a weakly hydrophobic surface, our droplets do not have a proper ellipsoidal shape, hence global factors are likely to be introduced in all geometrical relations used to derive the potential (\ref{ellipsoidal_droplet_pot}), e.g. equations (\ref{averaged_N}), (\ref{geometrical_relations}), (\ref{f_geometrical_function}).

Taking into account both factors in the resolution, we obtain the results plotted on the left in figure \ref{fig:Figure_13} for the magnets used by \citet{Timonen2013} and on the right for our magnets, whose characteristics are detailed in section \ref{sect:magnets}. The gradient squared is taken into account in $B_\nabla$, and the variation of the demagnetization factor with the droplet shape is taken into account in $B_\lambda$. The neglected quantities are therefore the horizontal gradient, the higher order derivatives, and the inhomogeneity of the demagnetization.

The eight behaviours from the plots are similar, and allow us to observe four steps. 
First: the aspect ratio increases. Second: after reaching its maximum, it starts decreasing. Third: the fission occurs. Fourth: other fissions occur while the aspect ratio converges towards a constant value around one. This process perfectly describes the experimental observation (see movie S1 of \citet{Timonen2013}).
It can be understood in a qualitative way: first the gradient is negligible, hence the shape is driven by the demagnetization which increases the aspect ratio, exactly as observed by \citet{Bacri1982,Afkhami2010,Zhu2011}. The relevant diagram is the one plotted in figure \ref{fig:Figure_7}, with a curve of the blue one type ($\chi \sim 5.4$). Then the gradient becomes non negligible and its contribution opposes the demagnetization: the aspect ratio goes backwards and the droplet flattens. The relevant diagram showing this decrease is the one plotted in figure \ref{fig:Figure_9}. Then the gradient becomes so high that it is energetically favorable to create several droplets, resulting in an increased surface but smaller height. The aspect ratio then quickly reaches 1 after the first fissions occurred. Eventually, the aspect ratio is fixed to 1 and the number of droplets converges to the law $n \propto \lambda^{-3}$, with, as explained above, the same slope in our case (the squared gradient correction is negligible) but different slopes in the Timonen \textit{et al.} case (the squared gradient correction is not negligible and induces a hierarchy). Note that the ``Flat Neodymium'' magnet is not strong enough to break the droplet after the first fission. 
%

\section{Experimental setup}
\label{sect:exp_setup}
In this section, after a characterization of our experimental setup, we  aim to experimentally check the validity of the fissioning behavior, derived from our theoretical energetic study, and its dependence on the external parameters such as the fluid's density, the droplet's initial volume, and the magnetic gradient. To do so, we essentially focus on the first splitting of an aqueous ferrofluid droplet subjected to a magnetic field generated by various permanent magnets.

\subsection{Characteristics of the ferrofluid}
\label{characteristics}
%
\begin{figure}
  \centerline{\includegraphics[width=0.6\linewidth]{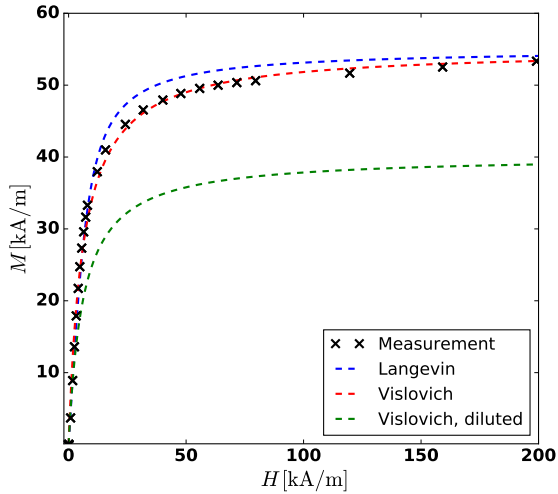}}
\caption{Crosses: experimental magnetization curve. Dotted lines: fit obtained with equation (\ref{arctan}), and $M_s$ and $\chi$ as fitting parameters.}
\label{fig:Figure_14}
\end{figure}
%
The experiments are carried out with an aqueous ionic ferrofluid, with Cobalt ferrite particles ($\mathrm{CoFe}_2\mathrm{O}_4$) with a mean size of about 10 nm \citep[see][]{Neveu1993}, of density $\rho = 2150 \pm 10 \text{ kg m$^{-3}$}$ and surface tension $\sigma = 50 \pm 2 \text{ mN/m}$, provided by Sophie Neveu (Universit\'e Pierre et Marie Curie, CNRS) and Florence Elias (Laboratoire Mati\`ere et Syst\`e\-mes Complexes, Universit\'e Paris Diderot, CNRS). We repeated the experiments with its 2 times diluted version, of density $\rho = 1570 \pm 10 \text{ kg m$^{-3}$}$ and surface tension $\sigma = 65 \pm 2 \text{ mN/m}$.

The magnetization curve has been measured by the ferrofluid manufacturer using a vibrating sample magnetometer. It is given in figure \ref{fig:Figure_14}.
The magnetic susceptibility and saturated magnetization for the non diluted ferrofluid are $\chi = 5.4$ and $M_s = 55 \text{ kA/m}$.

The Langevin equation \citep[e.g.][]{Odenbach2009b}:
\begin{equation*}
M(H) = M_s \left( \coth \alpha - 1/\alpha \right), \quad \alpha = \mu_0 m H / (k_B T) = 3 \chi H /M_s
\end{equation*}
with $M_s$ the saturated magnetization, $H$ the magnetic field, $m$ the magnetic moment of a particle, $k_B T$ the thermal energy of a particle, and $\chi$ the initial susceptibility,
describes only approximatively the magnetization curves of our original ferrofluid and its diluted version. This well-known discrepancy is mostly attributed to the polydispersity of the fluid, and also to the interactions between the particles in concentrated fluids \citep{Shliomis1974,Berkovsky1993}, even if the latter can be taken into account in more sophisticated models \citep{Ivanov2007}.
We chose consequently to fit our data, as plotted in figure \ref{fig:Figure_14}, using the Vislovich approximation \citep{Vislovich1990}:
\begin{equation}
M(H) = \frac{M_s H}{H + H_T}
\label{arctan}
\end{equation}
where $H_T$ is the magnetic field strength such that $M(H_T) = M_s/2$. The magnetization curve of the diluted ferrofluid is obtained by multiplying the one of the non-diluted ferrofluid by the ratio of the densities (i.e. a factor 0.73).
Relation (\ref{arctan}) is used to determine the magnetization $M(H_a(0))$ and the differential susceptibility $\tilde{\chi}$ in the theoretical expressions.

\subsection{Magnets}
\label{sect:magnets}
\begin{figure}
\centerline{
\hspace{0.2cm}
\resizebox{0.55\textwidth}{!}{\includegraphics{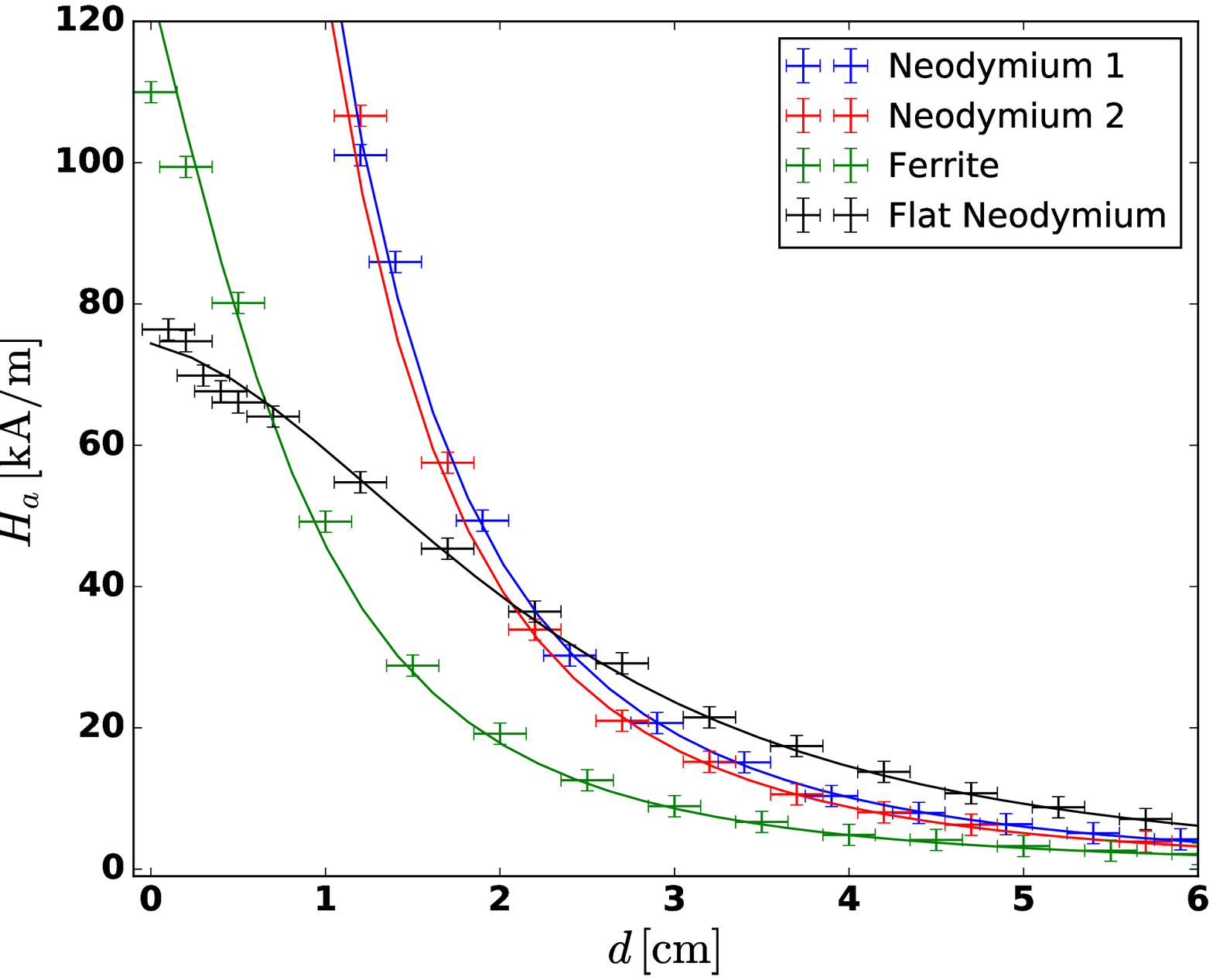}}
\hspace{-0.75cm}
\resizebox{0.6\textwidth}{!}{\includegraphics{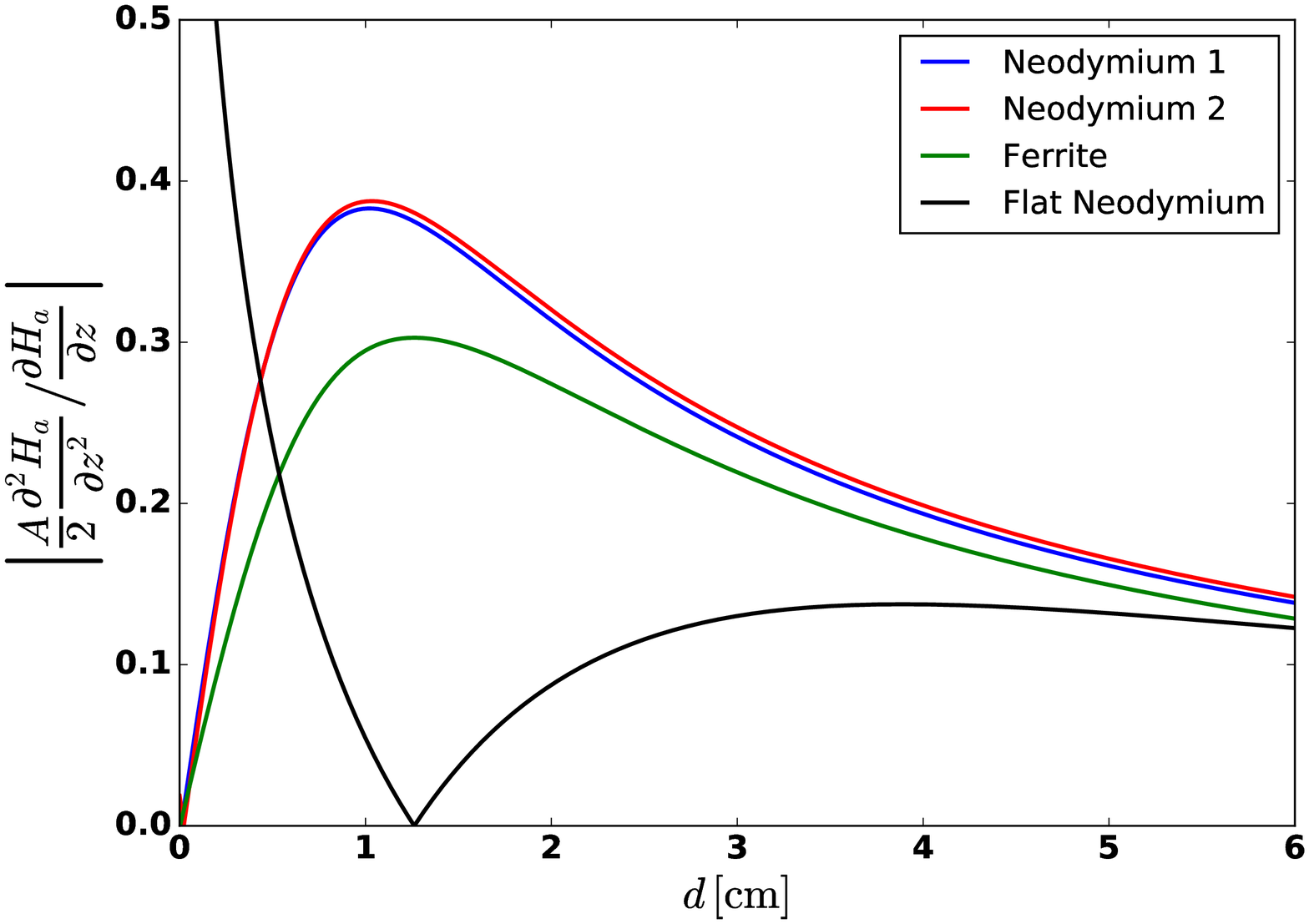}}}
\caption{Left: magnetic field strength of our four magnets as a function of the distance from their surface on the vertical axis. 
Right: ratio of the second derivative of the magnetic field over the gradient, for a typical droplet height $A=5$ mm.}
\label{fig:Figure_15}
\end{figure}

Our experiments were carried out using four different types of axisymmetric permanent magnets: two piles of neody\-mium magnets, labeled ``Neodymium 1'' and ``Neodymium 2'', one pile of ferrite magnets labeled ``Ferrite'' and one large flat neodymium magnet labeled ``Flat neodymium''. Their respective dimensions are listed in table \ref{tab:Table_2}. Their calibrations are shown in the left plot of figure \ref{fig:Figure_15}. The values of the magnetic field are measured using a Hirst model GM07 gaussmeter, for the field on the $z$ axis as a function of the distance $d$ to the magnet. The fits are obtained using the  analytic expression of the magnetic field created by a cylindrical magnet \citep[e.g.][]{Camacho2013}, which is analogous to the magnetic field created by a finite solenoid:
\begin{equation}
H_a = \frac{M_0}{2} \left( \frac{d+h}{\sqrt{ (d+h)^{2}+R^2 }} - \frac{d}{\sqrt{ d^{2}+R^2}} \right)
\label{formule_calibration}
\end{equation}
with $R$ the magnet radius, $h$ the magnet height and $M_0$ the intrinsic magnetization to be adjusted. The order of magnitudes for the latter are tabulated: $M_0 \sim 950$ kA/m for the neodymium magnets and $M_0 \sim 200$ kA/m for the ferrite magnet. The values used to obtain the fits are indicated in table \ref{tab:Table_2}.

We can infer from this plot a clear difference between the behaviors of the ``Flat neodymium'' magnet and the other magnets. In the case of the ``Flat neodymium'', the gradient is not a monotonous function of $d$: its maximum value is reached at $d = 1.15$ cm $\neq 0$, and it is almost vanishing at $d=0$. This difference in the generated magnetic field predicts, starting from the same theoretical expressions, the observation of clear discrepancies in the behaviour for small values of $d$ between the different magnets. These discrepancies should play the role of indicators to check the robustness of our theory in the experiments.

For our magnets, we typically have:
\begin{gather}
\frac{A}{2}
\frac{\mathrm{d}^2 M}{\mathrm{d}z^2} \sim 0.1 \, \frac{\mathrm{d} M}{\mathrm{d}z} \notag \\
N \tilde{\chi} \sim 0.1 \notag \\
\rho g \sim 0.2 \, \tilde{F_M} \notag \\
\frac{A}{2} B_\nabla  \sim 0.5 B_\lambda
\label{approx_exp}
\end{gather}
where $A \sim 5$ mm is the typical height of the droplets.
The right plot of figure \ref{fig:Figure_15} shows the ratio of the second derivative and the gradient for a typical height of 5 mm. The second derivative is in general small but non negligible: it will introduce an observable correction. It becomes negative for the ``Flat Neodymium'' magnet for $d\lesssim 1.2$ cm (this is why, in absolute value, the derivative is discontinuous at $d\simeq 1.2$ cm).
\begin{table}
  \begin{center}
\def~{\hphantom{0}}
  \begin{tabular}{lccc}
      \bf{Type} & $R$ [cm] & $h$ [cm] & $M_0$ [kA/m] \\[3pt]
        Neodymium 1    & 1    & 3.1 & 1011 \\
	    Neodymium 2    & 1    & 2.3 & 1003 \\
		Ferrite 	   & 1.25 & 6   & 255  \\
		Flat neodymium & 3    & 0.5 & 923  \\
  \end{tabular}
  \caption{Characteristics of the magnets. $R$: radius, $h$: width, $M_0$: intrinsic magnetization.}
  \label{tab:Table_2}
  \end{center}
\end{table}

%
\subsection{Overall procedure}
\label{procedure}
We determined the droplet's volume thanks to a Denver Instruments Pinnacle scale, model PI-403, sensitive to 1 mg. We estimated the uncertainty resulting of manipulation errors to be of about 3 mg. We then placed the droplet on a horizontal support, vertically aligned with one of our four magnets placed underneath. The support was maintained using a metal arm and its horizontality was checked using a spirit level. We slowly raised the magnet using a horizontal lifting plate, being careful to stay as much as possible in the quasistatic regime, until we observed the fission of the droplet, i.e. two separated droplets like in figure \ref{fig:Figure_6} (c). Note that in figure \ref{fig:Figure_6} (b), two non-separated peaks are observed but this is only due to the large amount of fluid required to get a picture of the transition. In practice, the splitting occurs quasi instantaneously: when reaching the splitting instability, an infinitesimal displacement of the approaching magnet is sufficient to spawn a clearly visible gap between the two daughter droplets (as in figure \ref{fig:Figure_6} (c)). Note that in our experiment this “infinitesimal” displacement is always smaller than the uncertainty on the measure of the distance to the magnet (1~mm), so that there is no ambiguity on the fission criterion.

We can then obtain the distance between the magnet and the fluid at first fission for a given droplet mass, with a precision of 1 mm. We carried out this experiment for each magnet with both ferrofluids (diluted and non-diluted), with the exception of the ``Flat neodymium'' magnet in the case of the diluted ferrofluid. In this case, the magnet needed to be approached very close, and the vertical magnetic force in the droplet was too weak compared to the horizontal magnetic field gradient. The droplet was observed to move toward the edge of the magnet and split earlier than expected, leading to a huge discrepancy in comparison with all the other cases, rendering its study irrelevant.

Our surface, made out of a thin polystyrene plastic cup,
is not superhydrophobic. Yet, it is hydrophobic enough for the splitting of the droplets to be clearly observed, and the created droplets are completely separated, as seen in figure \ref{fig:Figure_6} (except figure \ref{fig:Figure_6} (b), which is not representative of our experiments). The additional surface tension term in the potential is negligible and we will therefore neglect this issue while carrying out the experiments.

In the following, we focus the study on the first fission, measuring the distance between the magnet and the fluid at first splitting for different initial droplets' volumes.

%
%
\section{Results}
\label{results}
%
%
\subsection{Influence of the external magnetic field and concentration}
\begin{figure}
  \centerline{\includegraphics[width=1.2\textwidth]{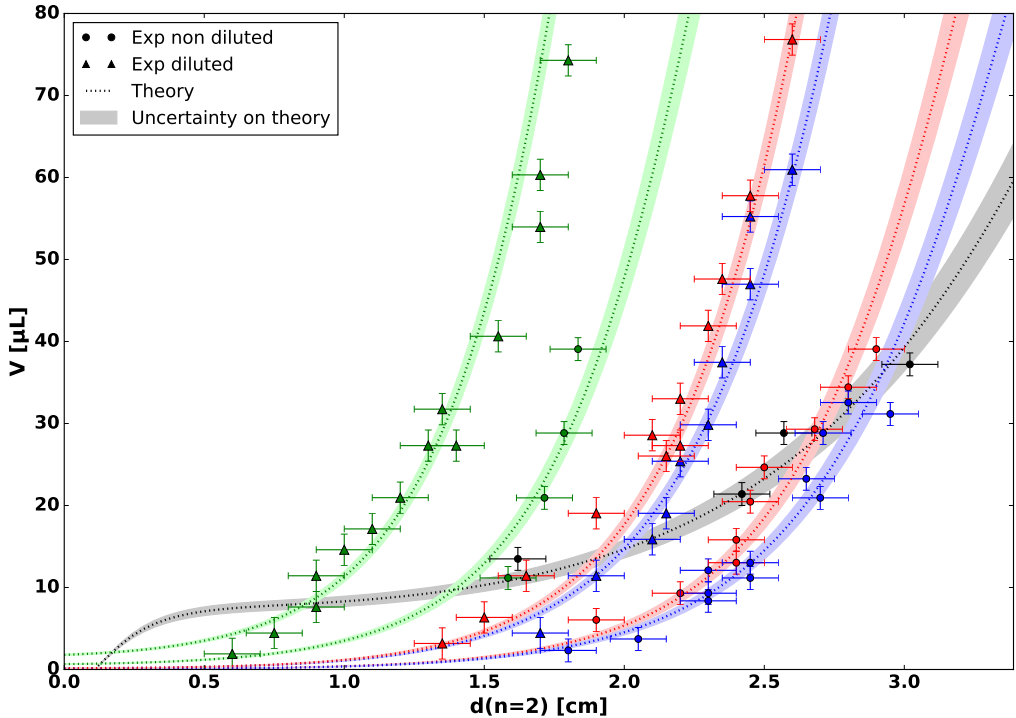}}
  \caption{Initial droplet's volume as a function of the distance from the magnet at first fission, compared between the magnets and between the two concentrations of ferrofluid. Each color corresponds to a specific magnet: blue: neodymium 1, red: neodymium 2, green: ferrite, black: flat neodymium. Circles: original non-diluted ferrofluid, triangles: diluted ferrofluid. The theoretical curves are obtained by numerically inverting equation (\ref{number_droplets}) and introducing a global factor of 1.35, while taking into account the squared gradient and second magnetic field derivative into the term $B_\nabla$.}
  \label{fig:Figure_16}
\end{figure}
%
We start by plotting the raw data, i.e. the initial volume of the droplet as a function of the distance from the magnet when the droplet splits in two daughter droplets (``first fission''). The results are shown in figure \ref{fig:Figure_16}. 
The blue, red, green and black colors correspond respectively to the magnets ``Neod\-ymium 1'', ``Neodymium 2'', ``Ferrite'', ``Flat neody\-mium''. The colored areas correspond to the uncertainties on the surface tension and fluid density, which are parameters of the theoretical expressions.
The theoretical curves are obtained by fixing $n=2$ in equation (\ref{number_droplets}) (taking into account the squared gradient and second magnetic field derivative into $B_\nabla$) and inverting it numerically to get $V$ as function of the external parameters. 

As previously explained, a global factor of 1.35 has also been added in front of $\mathcal{C}(K_c)$ to fit the experiments, and is attributed to the modified geometry of our droplets, which affects the expression of the demagnetization factor (equation (\ref{averaged_N})), as well as equations (\ref{ellipsoidal_droplet_pot}), (\ref{geometrical_relations}), (\ref{potential_aspect_ratio_1}), (\ref{f_geometrical_function}),... In any case, this factor is universal for a given geometry, as seen in our experiments: it does not depend neither on the magnets nor on the properties of the ferrofluid, and only needs to be calibrated once and for all.

$M(H_a(0))$ is determined from equation (\ref{arctan}) fitted by the measurements of figure \ref{fig:Figure_14}. $\left. \frac{\mathrm{d} H_a }{\mathrm{d}z} \right|_0$ is determined from equation (\ref{formule_calibration}) fitted by the measurements of figure \ref{fig:Figure_15}.
We can see a very good agreement between the theory and the experiments.
As expected, the stronger the external magnetic field gradient, the easier the fission. It is also harder to split diluted droplets (which get less easily magnetized): for the same initial mass, we need to approach the magnet closer.

The collapse of the black curve to zero is caused by the second derivative of the magnetic field, which becomes negative below $d\simeq 1$ cm and dominant below $d\simeq 0.5$ (see figure \ref{fig:Figure_15}) and makes the fission easier, until reaching a point where the potential is not bounded from below anymore, making $n$ to diverge to infinity for any initial volume. The theoretical prediction therefore breaks down below this distance (only for the ``Flat Neodymium'' magnet) as the higher order terms become no longer negligible.

%
\subsection{Universal behaviour}
\label{universal_behaviour}
\begin{figure}
  \centerline{\includegraphics[width=1\linewidth]{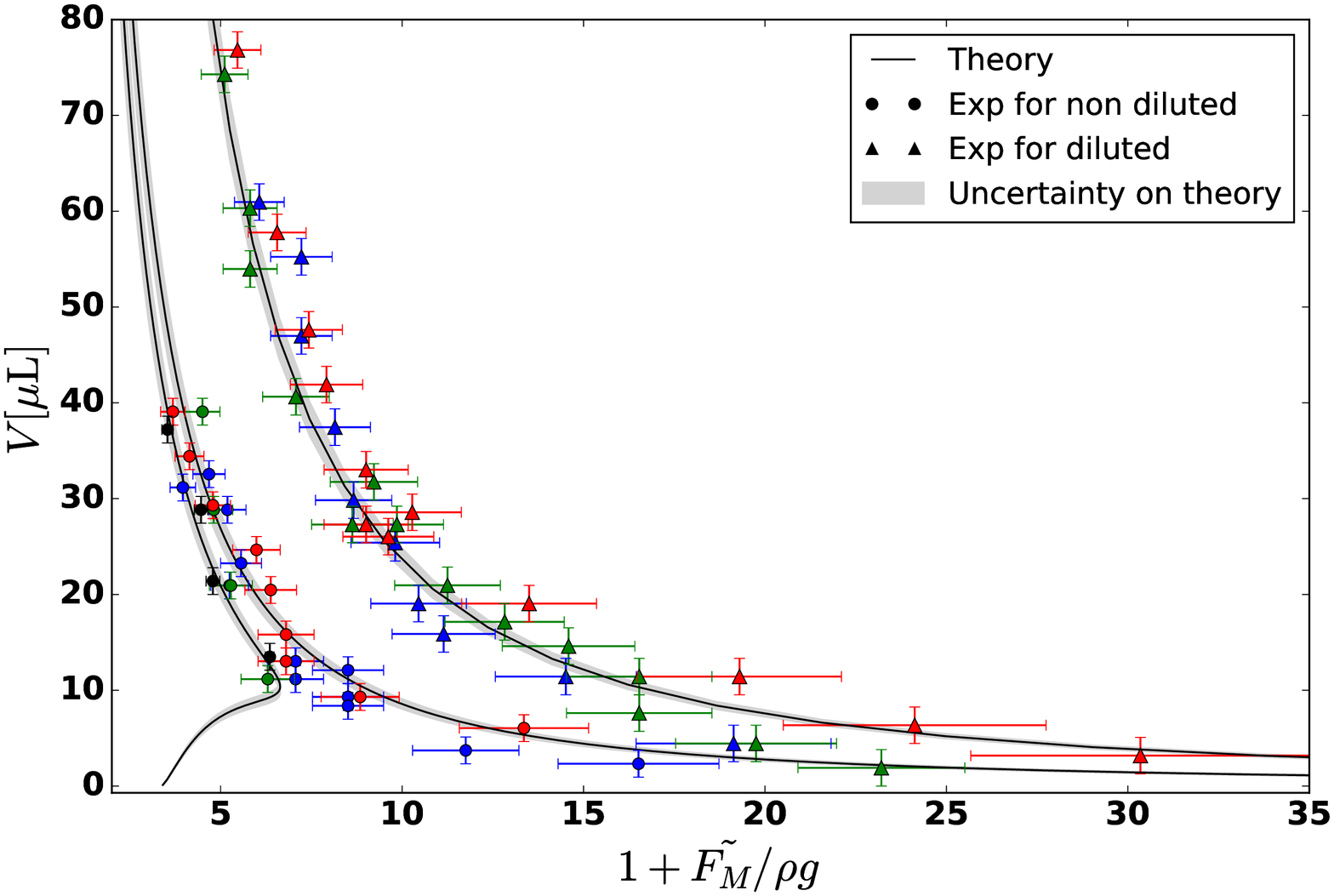}}
  \centerline{\includegraphics[width=1\linewidth]{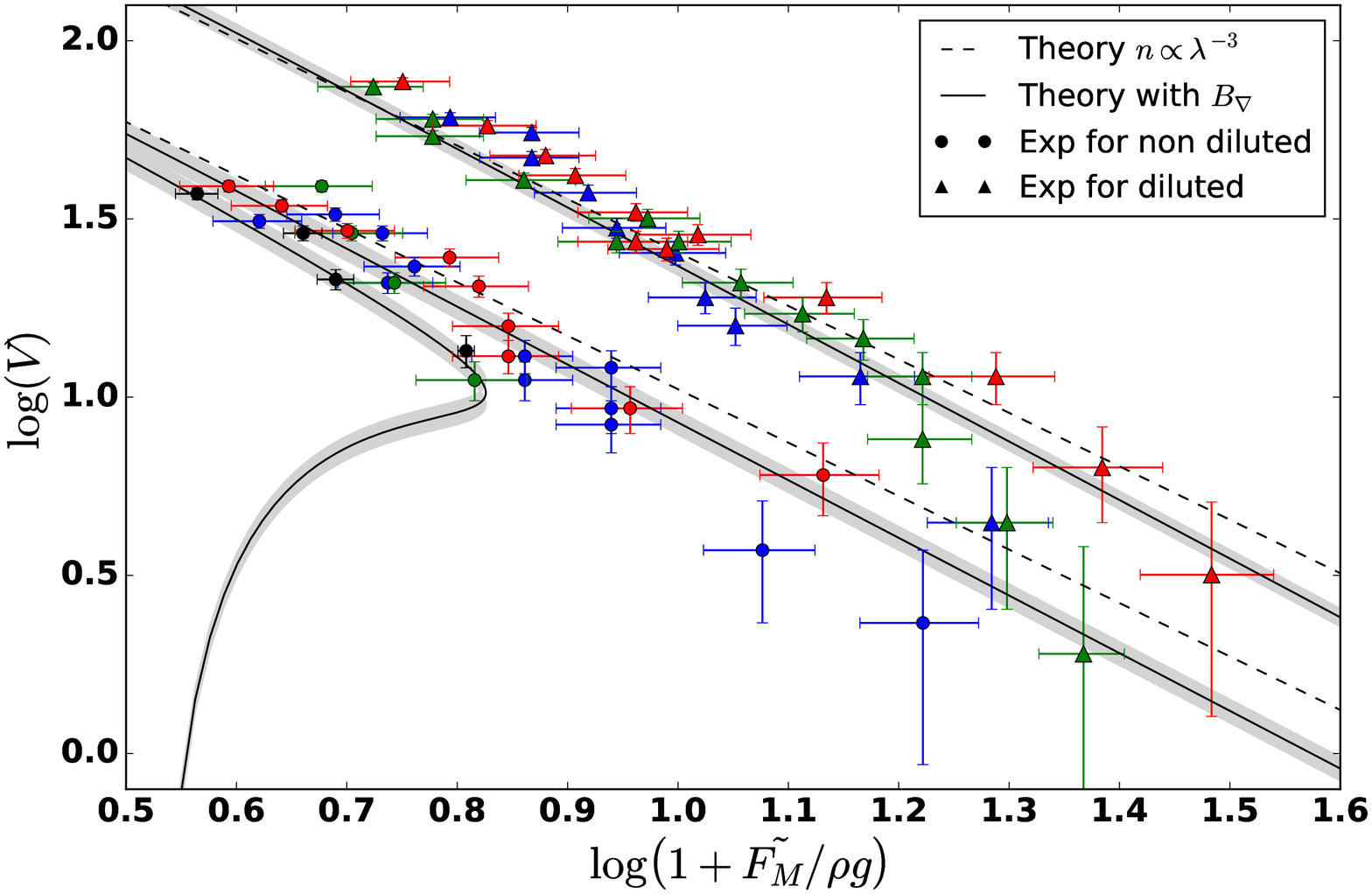}}
\caption{Universal behaviour for the first fission of a ferrofluid droplet, obtained by plotting the initial volume of the droplet as a function of the external field applied at first fission. The colors and symbols are identical to the ones of figure \ref{fig:Figure_16}.
The dashed lines show the behaviour predicted by the proportionality law parametrized with a global factor 2.7. The solid lines and grey areas are theoretical predictions obtained by numerically solving equation (\ref{number_droplets}) (taking into account the squared gradient and second derivative) and introducing a global factor of 1.35. The hysteresis is the theoretical prediction obtained for the ``Flat Neodymium'' magnet when taking into account the squared gradient and second derivative but not the higher order terms.}
\label{fig:Figure_17}
\end{figure}

In order to compare the behaviours for the different magnets, we plot the initial volume of the droplets as a function of the magnetic force $(1+ \tilde{F_M}/\rho g)$ that acts on them when the first fission ($n=2$) occurs. The result is plotted in figure \ref{fig:Figure_17}. Such plot erases the differences observed between the magnets in figure \ref{fig:Figure_16}. 
For a given ferrofluid, all experimental points fall on the same straight line on the log-log plot. This verifies again that in our case the slopes of equation (\ref{number_droplets}) are similar for all magnets, and that the aspect ratio at first fission is roughly the same for all initial volumes, dilution, and magnets.
The theoretical curve for the ``Flat neodymium'' magnet shows an hysteresis for high values of $\tilde{F_M}$ because of the second magnetic field derivative contribution. Note that there is no experimental point beyond the theoretical breakdown.

Taking only into account the simple proportionality law $n \propto \lambda^{-3}$ (which here essentially amounts to neglecting $B_\nabla$), log($V$) should scale as $(1 + \tilde{F_M}/\rho g )^{-3/2}$ whatever the magnet. This corresponds to the dotted lines in figure \ref{fig:Figure_17} (parametrized with a global factor 2.7). We observe a good correlation when the magnetic force is weak. However, a discrepancy is observed for higher magnetic forces: a numerical fit of the experimental data gives a slope of -1.8 for both dilutions. The proportionality law overestimates the magnetic force needed to fission the droplet. This is because the second derivative of the magnetic field has to be taken into account when dealing with high magnetic fields (i.e. small distances between the magnet and the droplet ($\sim$ 1 cm, see figure \ref{fig:Figure_15})). Taking into account $B_\nabla$ by numerically inverting equation (\ref{number_droplets}) leads to a very good correlation between the theoretical expectation and the experimental behavior. The correction is non trivial: since $B_\nabla$ counterbalances the gradient, we should expect to need a higher $\tilde{F_M}$ in order to fission a given volume, contrary to what is observed. This is because the second derivative contribution itself depends on the volume through the $R_0^3$ factor in equation (\ref{potential_n_droplets}). If the volume is huge, then the droplet is \textit{a priori} easier to fission, but the second derivative contribution is also higher, then the droplet is also harder to fission. This is why, in the end, smaller droplets are easier to fission when taking this term into account: the second derivative contribution has less impact. Of course, all volumes should globally be harder to fission, hence the dashed lines should be below the solid lines. In fact, this is indeed the case if the same global factor is taken for both lines. In the plot of figure \ref{fig:Figure_17}, the factor 1.35 correcting the solid line is indeed smaller than the factor 2.7 correcting the dashed line (and closer to 1).
%

%
\subsection{Minimum reachable droplets' volume}
\label{minimum_volume}
Figure \ref{fig:Figure_16} predicts that it may happen for the initial droplet's volume to be too small for a fission to occur: apart from the ``Flat Neodymium'' case, the theoretical curves do not converge to zero. We seek here to experimentally check this and further investigate the behaviour of the ``Flat Neodymium'' curve at high field, by directly determining the critical size below which the fission no longer occurs, even when the magnet is in contact with the ferrofluid ($d=0$). To experimentally evaluate this limit size, we created droplets of decreasing size and tried to fission them.

We keep as the limit value the mass of the biggest droplet for which the fission was not observed, even at zero distance. As expected, in the case of the ``Flat Neodymium'' magnet, the theory breaks down: we experimentally evidenced the existence of a finite minimum volume which is not predicted on figure \ref{fig:Figure_16}. This minimum volume ($\sim 10$ $\umu$L) corresponds in fact to the plateau (where the theory is still valid), between 0.5 and 1 cm.
For the other three magnets, in order to compare with the theory, we numerically solve simultaneously equation (\ref{potentiel_minimization_K}) and equation (\ref{number_droplets}), with the field configurations at $d=0$ (highest $B_\lambda$ and lowest $B_\nabla$) as input. Focusing on this single configuration allows a better numerical precision and slightly change the values which can be directly read on figure \ref{fig:Figure_16} (of $\sim$ 1 $\umu$L).

\begin{table}
  \begin{center}
\def~{\hphantom{0}}
  \begin{tabular}{ccccc}
      Magnet & Neodymium 1 & Neodymium 2 & Ferrite & Flat neodymium \\[3pt]
        $V_{c,theo}$ [$\umu$L] & $[0.18,0.27]$ & $[0.20,0.31]$ & $[1.2,1.4]$ & $[6.6,8.9]$ \\
		$V_{c,exp}$ [$\umu$L] & $< 0.4$ & $\lesssim 0.4$ & $[1.2,1.6]$ & [7.7,8.3] \\
  \end{tabular}
  \caption{Results for the minimal volume of droplet allowing the splitting phenomenon.}
  \label{tab:Table_3}
  \end{center}
\end{table}

Theoretical and experimental results for the droplets' minimal volumes are summarized in table \ref{tab:Table_3}. The theoretical interval for the ``Flat Neodymium'' magnet is obtained by taking the theoretical minimum value at $d=0.5$ cm and the theoretical maximum value at $d=1$ cm.
The uncertainty on the other theoretical values comes from the uncertainty on the surface tension and density.
The experiments were repeated several times to decrease the statistical uncertainty. The minimal volume is of the order of the tenth of microliter for our most powerful neodymium magnets, of the microliter for our ferrite magnet, and of the order of ten microliters for our flat neodymium magnet. The agreement between the theoretical and experimental values is very good for the ``Ferrite'' magnet, and good for the ``Neodymium 2'' magnet, although we reach the limits of our experimental precision. In the case of the ``Flat neodymium'' magnet, the agreement is very good taking the plateau for the theoretical computation.

%
\section{Conclusion}
\label{conclusion}
We studied the behaviour of a ferrofluid distribution placed in a non-uniform magnetic field.
We first derived a general potential describing a volume of ferrofluid placed in a vertical magnetic field gradient, without assuming a specific regime for the magnetization and not limiting ourselves to the magnetic gradient.

We applied this formalism to specific geometries, first showing how the normal-field instability of a layer of fluid is modified in presence of a weak gradient, second studying the evolution of a ferrofluid droplet.
The instability phenomenon observed in the case of a thin layer of fluid and those observed in the case of a ferrofluid droplet appear as equivalent, which is quite surprising since we studied the layer within a linear analysis and the droplet within a geometrical formalism describing a highly curved interface.
This shows that a similar origin underlies all those experimental observations.

Focusing on a single droplet, we derived expressions allowing to predict the evolution of the aspect ratio and showed a good agreement with the experiments of \citet{Timonen2013}, without any free parameter.
We also predicted new aspect ratio instabilities showing up because of the competition between the uniform field contribution, the magnetic field gradient, and the surface tension. Those instabilities should be observable through discontinuities in the evolution of the aspect ratio. In particular we evidenced a prolate-oblate transition.

We then investigated in details the origin of the fission phenomenon observed by \citet{Timonen2013}, essentially driven by the competition between the magnetic field gradient and the surface tension. We proved the proportionality law $n\propto \lambda^{-3}$ guessed by \citet{Timonen2013}, which appears as the first approximation of a more general expression given by equation (\ref{number_droplets}). New contributions like the squared gradient and the second magnetic field derivative introduce small but non negligible corrections which satisfactorily explain the slope hierarchy observed in \citet{Timonen2013} and, up to a small global correction, lead to a very good agreement with our own experiments, focused on the first fission.

A numerical resolution allows to predict the evolution of both the number of droplets and their aspect ratio. The result describes the observed phenomenon: the droplet stretches, then flattens a bit, and finally fissions.

The minimal volume of droplet allowing the fission is also in agreement with the theory, and is of the order of 0.1 $\umu$L for our most powerful magnets, i.e. three orders of magnitude below the minimum volume allowing the fission under an uniform magnetic field.
It is then theoretically possible to use this topological instability in order to create nanodroplets of ferrofluid using magnetic fields of the order of 0.1 T.

Our work proved that a ferrofluid in a non-uniform magnetic field has a very rich phenomenology which can be analytically studied. We aimed to understand the fundamental mechanisms leading to various experimental observations. We discussed previous and current experiments, and suggested new possibilities of studies. The predicted phase transitions in the aspect ratio evolution should be observable. It would also be interesting to investigate the effects of the horizontal gradient of a magnetic field, which can in particular induce asymmetries in the fluid distribution. Finally, the understanding of the fission mechanism may have applications for the creation and manipulation of nanodroplets of precise size, using combinations of uniform fields and gradients.

\bigbreak \bigbreak
We gratefully acknowledge Dr.~Florence Elias (Laboratoire Mati\`ere et Syst\`emes Complexes, Universit\'e Paris Diderot, CNRS) and Dr.~Sophie Neveu (Universit\'e Pierre et Marie Curie, CNRS) for providing us with aqueous ferrofluid. We also thank Florence Elias for helpful discussions and suggestions on the manuscript.

We are more than grateful to Claire Marrache-Kikuchi, Fr\'ed\'eric Bouquet, Bertrand Lacroix-A-Chez-Toine, Jean-Marie Fischbach and Bertrand Pilette for their great commitment and the time they spent guiding and supporting us. In particular, we thank Claire Marrache and Fr\'ed\'eric Bouquet for very helpful discussions and valuable corrections on the manuscript.

We also thank the Paris-Sud University and Fundamental Physics Magist\`ere for supporting us and providing us with the tools and materials required for this work.

We finally thank the organization of the French Physicists' Tournament\footnote{http://france.iptnet.info/} for giving us the inspiration to work on this fascinating subject, as well as the other members of the Paris-Sud team: Luc Barast,  Lydia Chabane, Sarah Christin, Daniela Galarraga, Thibaut Perdereau and Guillaume Thiam, for their daily encouragements and for the patience they showed during our endless discussions.

%

\appendix
\section{Demagnetization factor in non-uniform magnetic fields}
\label{appA}
\subsection{General formalism}
Let us consider a sample magnetized under the influence of an external field $\boldsymbol{H_a}$. The total field $\boldsymbol{H}$ inside the sample can be written as the superposition of the induced field $\boldsymbol{H_d}$ and the applied field:
\begin{equation}
\boldsymbol{H} = \boldsymbol{H_a} + \boldsymbol{H_d}
\end{equation}

The field $\boldsymbol{H_d}$ results from the interactions between the magnetized dipoles, which create a volume current density $\boldsymbol{\nabla} \times \boldsymbol{M}$ inside the sample and a surface current $\boldsymbol{M} \times \boldsymbol{n}$ at the surface of the sample.

Maxwell's equations imply that $\boldsymbol{\nabla} \times \boldsymbol{H_d} = 0$, which means that the induced field can be rewritten as the gradient of a scalar field $\Phi$:
\begin{equation}
\boldsymbol{H_d} = -\boldsymbol{\nabla} \Phi(\boldsymbol{r})
\end{equation}
On the other hand, we have $\boldsymbol{\nabla} \cdot \boldsymbol{H_d} = - \boldsymbol{\nabla} \cdot \boldsymbol{M}$, so that the scalar field has to obey:
\begin{equation}
\nabla^2 \Phi = \boldsymbol{\nabla} \cdot \boldsymbol{M}
\end{equation}
which is the Poisson equation for a distribution of charges $-\boldsymbol{\nabla} \cdot \boldsymbol{M}$. The general solution can be written in terms of a Green function as:
\begin{equation}
\Phi(\boldsymbol{r}) = -\frac{1}{4 \pi} \int_V \frac{\boldsymbol{\nabla'} \cdot \boldsymbol{M}(\boldsymbol{r'})}{\| \boldsymbol{r}-\boldsymbol{r'} \|} dV'
\end{equation}
which can be rewritten as:
\begin{equation}
\Phi(\boldsymbol{r}) = -\frac{1}{4 \pi} \int_V \boldsymbol{M}(\boldsymbol{r'}) \cdot \boldsymbol{\nabla'} \left[ \frac{1}{\| \boldsymbol{r}-\boldsymbol{r'} \|} \right] dV'
\end{equation}
which is the expression given by \citet{Newell1993}.

Therefore, the induced field, or demagnetizing field, is given by:
\begin{equation}
\boldsymbol{H_d} = - \frac{1}{4 \pi} \boldsymbol{\nabla} \int_V \boldsymbol{M}(\boldsymbol{r'}) \cdot \boldsymbol{\nabla'} \left[ \frac{1}{\| \boldsymbol{r}-\boldsymbol{r'} \|} \right] dV'
\label{demagnetization_field}
\end{equation}
In the case of an uniform magnetization (for uniformly magnetized ellipsoids), equation (\ref{demagnetization_field}) reduces to a tensorial relation, written with implicit summation as:
\begin{equation}
{H_d}_{i} = - M_i N_{ij}
\label{def_N}
\end{equation}
the demagnetization tensor $N_{ij}$ being defined by:
\begin{equation}
N_{ij} \equiv - \frac{1}{4 \pi} \int_V \partial_i \partial_j \left[ \frac{1}{\| \boldsymbol{r}-\boldsymbol{r'} \|} \right] dV'
\end{equation}
If the sample has an ellipsoidal shape and is uniformly magnetized, the tensor is diagonal in the basis given by the three principal axis, hence for an ellipsoid magnetized along one of the principal axis, the tensor reduces to a scalar, and we obtain equation (\ref{self_consistent}).

\subsection{Generalization to non-uniformly magnetized samples}
We return to the expression of the demagnetizing field given by equation (\ref{demagnetization_field}):
\begin{equation}
\boldsymbol{H_d} = -\boldsymbol{\nabla} \Phi(\boldsymbol{r}) = -\frac{1}{4 \pi} \boldsymbol{\nabla} \int_V \boldsymbol{M}(\boldsymbol{r'}) \cdot \boldsymbol{\nabla'} \left[ \frac{1}{\| \boldsymbol{r}-\boldsymbol{r'} \|} \right] dV'
\end{equation}
This has been derived in full generality, in particular it holds for samples in non-uniform magnetic fields.

Since $\boldsymbol{M}(\boldsymbol{r'})$ can not be extracted from the volume integral, the computation of $\boldsymbol{H_d}(\boldsymbol{r})$ at a given point requires to know the values of $\boldsymbol{M}$ at any point inside the sample. Importantly, this means that a local relation as:
\begin{equation}
\boldsymbol{H_d}_i(\boldsymbol{r}) = -N_{ij}(\boldsymbol{r}) \boldsymbol{M}_j(\boldsymbol{r})
\end{equation}
which is the naive generalization of equation (\ref{def_N}), cannot be exact. It would indeed imply a loss of information.

The idea is then to rewrite equation (\ref{demagnetization_field}) using the Taylor development of the magnetization at order $n \to \infty$, which amounts to defining an infinity of demagnetizing fields proportional to the derivatives of the magnetization, hence an infinite series of demagnetization factors, each of them defined through a volume integral.

We assume that the magnetization is aligned along the vertical axis and depends only on $z$: $\boldsymbol{M} = M(z) \boldsymbol{e_z}$. This is a reasonable assumption when considering an ellipsoidal sample magnetized under the action of an external field $\boldsymbol{H_a} = H_a(z) \boldsymbol{e_z}$, which is the system we will be considering from here on out. The origin of our coordinates is taken as the center of the ellipsoid and the axes are along the natural axes of the ellipsoid. The Taylor expansion can be performed at any point in the sample, since the knowledge of all derivatives of a given function at a single point is equivalent to the knowledge of the function at all points:
\begin{equation}
M(z') = \sum_{n=0}^\infty \frac{(z'-a)^n}{n!} \left. \frac{\partial^n M}{\partial z^n} \right|_a
\end{equation}
Note that $M(z')$ in $z'$ is completely encoded by the set of its derivatives in $a$, which means that information is conserved and the computation is consistent.

The demagnetizing field therefore reads:
\begin{equation}
H_d = -\sum_{n=0}^{\infty} N_n(a;\boldsymbol{r}) \left. \frac{\partial^{n} M}{\partial z^{n}} \right|_a
\label{demagnetizating_field_1}
\end{equation}
with $a$ a free parameter, and where we defined an infinite ``demagnetizing'' series of demagnetization factors:
\begin{align}
N_n(a;\boldsymbol{r}) \equiv - \frac{1}{4 \pi} \frac{1}{n!} \int_V 
\partial_z \left[(z'-a)^{n} \frac{(z'-z)}{\| \boldsymbol{r}-\boldsymbol{r'} \|^3} \right] dV'
\label{def_N_series}
\end{align}
Every term of the series is well defined through a volume integral, and can be computed either analytically or numerically.

\subsection{Properties of the demagnetizing series for ellipsoidal geometries}
Equation (\ref{demagnetizating_field_1}) is suitable for numerical simulations but not very convenient if one wants to get analytical results. The goal of this section is to investigate the relevance of the higher order terms in the expansion.

We have the following property:
\begin{align}
\frac{\partial N_{n+1}}{\partial a} = - N_n(a;z)
\label{eq_diff_N}
\end{align}
Hence the following recurrence relation:
\begin{equation}
N_{n+1}(a;\boldsymbol{r}) = - \int \mathrm{d}a N_{n}(a;\boldsymbol{r}) + f_{n+1}(\boldsymbol{r})
\end{equation}
If the magnetization is uniform, the demagnetizing field is:
\begin{equation}
H_d(\boldsymbol{r}) = -N_0(a;\boldsymbol{r}) M
\end{equation}
which means that, placing ourselves in the case of an ellipsoidal distribution, $N_0(a;\boldsymbol{r})$ is constant, and we define:
\begin{equation}
N_0(a;\boldsymbol{r}) \equiv N
\end{equation}
Equation (\ref{eq_diff_N}) therefore implies that the terms of the demagnetizing series are polynomial in $a$:
\begin{equation}
N_n(a;\boldsymbol{r}) = \sum_{k=0}^{n} \frac{(-a)^k}{k!} N_{n-k}(0;\boldsymbol{r})
\end{equation}

Since we can express every $N_{n}(a;\boldsymbol{r})$ in terms of $N_{k}(0;\boldsymbol{r})$ (with $k \leq n$), let us focus on $N_{n}(0;\boldsymbol{r})$, of which we recall the expression:
\begin{align}
N_n(0;\boldsymbol{r}) = - \frac{1}{4 \pi} \frac{1}{n!} \int_V 
\partial_z \left[(z')^{n} \frac{(z'-z)}{\| \boldsymbol{r}-\boldsymbol{r'} \|^3} \right] dV'
\label{def_N0_series}
\end{align}
Note that, in our case, the plane $(0,x,y)$ is a symmetry plane.
We therefore have:
\begin{equation}
N_n(0;(x,y,-z)) = (-1)^n N_n(0;(x,y,z))
\end{equation}
Let us consider the average of the demagnetizing field over the magnetized volume:
\begin{align}
\langle H_d(\boldsymbol{r}) \rangle_V &\equiv \frac{1}{V} \int_V dV H_d(\boldsymbol{r}) \\
&=
-\sum_{n=0}^{\infty} \left. \frac{\partial^{n} M}{\partial z^{n}} \right|_a \langle N_n(a;\boldsymbol{r}) \rangle_V
\end{align}

We simplify the notation as follows: $\langle N_n(a;\boldsymbol{r}) \rangle_V \equiv N_n(a)$. For $n$ odd, $N_n(0;(x,y,z))$ is an odd function in $z$, hence its integral over $z$ vanishes:
\begin{equation}
\forall n \text{ odd,  } \,\,\, N_n(0) = 0
\end{equation}
Hence:
\begin{gather}
\forall n \text{ even,  } \,\,\, N_{n}(a) = \sum_{k=0}^{n/2} \frac{a^{2k}}{(2k)!} N_{n-2k}(0)
\\
\forall n \text{ odd,  } \,\,\,  N_{n}(a) = -\sum_{k=0}^{(n-1)/2} \frac{a^{2k+1}}{(2k+1)!} N_{n-2k-1}(0)
\end{gather}

So that the demagnetizing field finally reads, in average:
\begin{equation}
\langle H_d \rangle_V = 
-M(a) N
+ \left. \frac{\partial M}{\partial z} \right|_a a N
- \left. \frac{\partial^{2} M}{\partial z^{2}} \right|_a N_2(0) - \left. \frac{\partial^{2} M}{\partial z^{2}} \right|_a \frac{a^2}{2} N
- \sum_{n=3}^{\infty} \left. \frac{\partial^{n} M}{\partial z^{n}} \right|_a N_n(a)
\end{equation}
In most cases, it is sufficient to reduce the analysis to the uniform contribution of the magnetization. The magnetization appears in equation (\ref{demagnetization_field}) as a function of the position: $\boldsymbol{M} = \boldsymbol{M}(\boldsymbol{r'})$, but in the case of a sample magnetized only under the influence of an external magnetic field, e.g. a paramagnetic fluid, the magnetization sees the position only through the applied field: $\boldsymbol{M} = \boldsymbol{M}(\boldsymbol{H_a}) = \boldsymbol{M}(\boldsymbol{H_a(\boldsymbol{r'})})$, and the $n$-derivative of $\boldsymbol{M}$ with respect to the position will be a term of order $n+1$ in the applied field, as given by the Fa\'a di Bruno formula.
For instance:
\begin{equation}
\frac{\partial^{2} M}{\partial z^{2}} = \frac{\partial^2 M}{\partial H^2} \left( \frac{\partial H}{\partial z} \right)^2  + \frac{\partial M}{\partial H} \frac{\partial^2 H}{\partial z^2}
\end{equation}
which is usually negligible. Let us therefore restrict the analysis to the first order:
\begin{align}
\langle H_d \rangle_V &\simeq -\left( M(a) - \left. \frac{\partial M}{\partial z} \right|_a a \right) N \\
&\simeq -N M(0)
\end{align}
which gives a precise meaning to the self-consistent equation (\ref{self_consistent}). It also shows that equation (\ref{self_consistent}) holds at first order when considering averages and taking the magnetization at the center of the geometry, and finally justifies why we can neglect the other terms, as they are made up of products of higher order derivatives of the magnetization and the applied field. Note also that one can consider the magnetization in another point in the geometry and neglect the correction of order $a$, generally small compared to the uniform contribution.

\section{Generalized linear approximation for the magnetization and higher order magnetic field derivatives}
\label{appB}
We start by recalling the expression of the magnetic volume potential from equation (\ref{surf_pots}), written with its full dependency for clarity:
\begin{equation}
\mathrm{d}^3 U_m = - \mu_0 \int_0^{H_a(z)} \mathrm{d} h_a M(H(h_a(z)))
\end{equation} 

The goal is to separate this expression into a uniform term, whose effects are known even outside the linear regime for the magnetization, and a non uniform term integrated on a small portion of the magnetization curve, so that we can approximate the magnetization by its tangent, generalizing the linear regime. We hence split the integral:

\begin{equation}
\mathrm{d}^3 U_m = - \mu_0 \int_0^{H_a(0)} \mathrm{d} h_a M(H)  - \mu_0 \int_{H_a(0)}^{H_a(z)} \mathrm{d} h_a M(H)
\label{integral_separation}
\end{equation}

Between $H_a(0)$ and $H_a(z)$, we can approximate the magnetization by its tangent, providing that for all $h_a$ between $H_a(0)$ and $H_a(z)$, $H$ is close to $h_a$, and that $H_a(z)$ and $H_a(0)$ are also close (if the droplet height is small). Note that this development is also valid if the second derivative of the magnetization is small between $H_a(0)$ and $H_a(z)$. In particular, it contains the linear regime and saturated regime approximations. In this sense, it is a generalization of the linear approximation defined in the origin to any linear portion of the magnetization curve.

\begin{align}
M(H(h_a)) &\simeq M(h_a) + (H(h_a)-h_a) \left. \frac{\partial M}{\partial H} \right|_{h_a} \\
&\simeq M(h_a) - N M(H(h_a)) \left. \frac{\partial M}{\partial H} \right|_{h_a} \notag \\
&\simeq M(h_a) \left( 1 + N \left. \frac{\partial M}{\partial H} \right|_{h_a} \right)^{-1} \notag
\end{align}
\begin{equation}
M(H(h_a)) \simeq \left(M(H_a(0)) + (h_a - H_a(0)) \left. \frac{\partial M}{\partial H} \right|_{H_a(0)} \right) \left(1 + N \left. \frac{\partial M}{\partial H} \right|_{h_a} \right)^{-1}
\end{equation}

We do the following approximation:
\begin{equation}
\forall h_a \in [H_a(0),H_a(z)] \,
\left. \frac{\partial M}{\partial H} \right|_{h_a} \simeq \left. \frac{\partial M}{\partial H} \right|_{H_a(0)} \equiv \tilde{\chi}
\label{generalized_susceptibility}
\end{equation}
which is in general only true on small portions of the magnetization curve, i.e. for $H_a$ close to $H_a(0)$, i.e. only suitable for the second integral in equation (\ref{integral_separation}). $\tilde{\chi}$ is the differential susceptibility.

The magnetization is therefore expressed by:
\begin{align}
M(H) &\simeq \frac{M(H_a(0))}{ 1 + N \tilde{\chi}} 
+ \frac{h_a \tilde{\chi}}{1 + N \tilde{\chi}} 
- \frac{H_a(0) \tilde{\chi}}{1 + N \tilde{\chi}}
\end{align}

In the usual linear approximation, $\tilde{\chi} = \chi$ and $M(H_a(0)) = \chi H_a(0)$, and we retrieve:
\begin{equation}
M(H) = \frac{H_a(z) \chi}{1 + N \chi} 
\end{equation}

In the saturated regime, $\tilde{\chi} = 0$ implies that the magnetization is indeed constant:
\begin{equation}
M(H) = M(H_a(0)) = M_s
\end{equation}

We can finally compute the second integral:
\begin{equation}
\mathrm{d}^3 U_m = - \mu_0 \int_0^{H_a(0)} \mathrm{d} h_a \, M(H)
 - \frac{\mu_0}{2}(H_a(z) - H_a(0)) \left( M(H(H_a(0)) + M(H) \right)
\label{general_magnetic_potential}
\end{equation}

In the linear regime we retrieve equation (\ref{mag_surf_pot}):
\begin{align}
\mathrm{d}^3 U_m &= - \mu_0 \int_0^{H_a(0)} \mathrm{d} h_a \, \frac{\chi H_a}{1 + N \chi} - \frac{\mu_0}{2}(H_a(z) - H_a(0)) \left( \frac{\chi H_a(0)}{1 + N \chi} + \frac{\chi H_a(z)}{1 + N \chi} \right) \notag \\
%
%
%
&= - \frac{\mu_0}{2} H_a(z) \chi H
\end{align}
%
%
%

In the saturated regime we retrieve the expected result:
\begin{equation}
\mathrm{d}^3 U_m = - \mu_0 H_a(z) M_s \notag
\end{equation}

Let us now write the magnetic surface potential with this generalized linear regime and taking into account all higher order derivatives of the magnetic field through its exact Taylor expansion in $z=0$:
\begin{equation}
H_a(z) = H_a(0) + \delta H_a (z)
\end{equation}
where
\begin{equation}
\delta H_a (z) = \sum_{n=1}^{\infty} \frac{z^n}{n!}
\left. \frac{\partial^n H_a}{\partial z^n} \right|_{0}
\end{equation}
and then injecting it into equation (\ref{general_magnetic_potential}), we get:
%
%
%
%
\begin{equation}
\mathrm{d}^3 U_m = 
- \mu_0 \int_0^{H_a(0)} \mathrm{d} h_a \, M(H)
- \frac{\mu_0}{2} \delta H_a(z) \frac{\tilde{\chi} \delta H_a(z)}{1 + N \tilde{\chi}}
- \frac{\mu_0}{1 + N \tilde{\chi}} \delta H_a(z) M(H_a(0))
\end{equation}

We can now integrate:
\begin{multline}
\mathrm{d}^2 U_m[h]
= - \mu_0 \int_0^h \mathrm{d}z \int_0^{H_a(0)} \mathrm{d} h_a \, M(H)
- \mu_0 M(H_a(0)) \sum_{n=1}^{\infty}  \left. \frac{\partial^n H_a}{\partial z^n} \right|_{0} \int_0^h \mathrm{d}z \, \frac{1}{1 + N(z) \tilde{\chi}} \frac{z^n}{n!} 
\\
- \frac{\mu_0}{2}
\sum_{i=1}^{\infty} \sum_{j=1}^{\infty}
\left. \frac{\partial^i H_a}{\partial z^i} \right|_{0} 
\left. \frac{\partial^j H_a}{\partial z^j} \right|_{0}
\int_0^h \mathrm{d}z \, \frac{z^{i+j}}{i!j!} \frac{\tilde{\chi}}{1 + N(z) \tilde{\chi}} \qquad \qquad
\end{multline}
%
%
where $N=N(z)$ is considered non-uniform in order to obtain general expressions also valid for non-ellipsoidal geometries.
Integrating by parts the second term leads to:
\begin{multline}
\mathrm{d}^2 U_m[h] = - \mu_0 \int_0^h \mathrm{d}z \int_0^{H_a(0)} \mathrm{d} h_a \, M(H) 
+ \frac{h^{2}}{2} \tilde{F_M}
- \mu_0 \sum_{n=2}^{\infty}  \left. \frac{\partial^n H_a}{\partial z^n} \right|_{0} 
\frac{M(H_a(0)) }{1 + N(h) \tilde{\chi}} \frac{h^{n+1}}{(n+1)!}
\\
- \mu_0 M(H_a(0)) \sum_{n=1}^{\infty}  \left. \frac{\partial^n H_a}{\partial z^n} \right|_{0}
\int_0^h \mathrm{d}z \, \frac{\tilde{\chi}}{(1 + N(z) \tilde{\chi})^2} \frac{z^{n+1}}{(n+1)!} \frac{\partial N}{\partial z}
\\
- \frac{\mu_0}{2}
\sum_{i=1}^{\infty} \sum_{j=1}^{\infty}
\left. \frac{\partial^i H_a}{\partial z^i} \right|_{0} 
\left. \frac{\partial^j H_a}{\partial z^j} \right|_{0}
\int_0^h \mathrm{d}z \, \frac{z^{i+j}}{i!j!} \frac{\tilde{\chi}}{1 + N(z) \tilde{\chi}} \qquad \qquad
\end{multline}
where
\begin{equation}
\tilde{F_M} \equiv - \mu_0 \frac{ M(H_a(0)) }{1 + N(h) \tilde{\chi}} \left. \frac{\partial H_a}{\partial z} \right|_{0}
\end{equation}

and the total surface potential reads:
\begin{multline}
\qquad \quad \mathrm{d}^2 U[h] = 
\frac{ h^2 }{2} \left( \rho g + \tilde{F_M} \right) 
+ \sigma \sqrt{1+ \left( \boldsymbol{\nabla} h\right) ^2 }
- \mu_0 \int_0^h \mathrm{d}z \int_0^{H_a(0)} \mathrm{d} h_a \, M(H) \\
- \frac{\mu_0}{2}
\sum_{i=1}^{\infty} \sum_{j=1}^{\infty}
\left. \frac{\partial^i H_a}{\partial z^i} \right|_{0} 
\left. \frac{\partial^j H_a}{\partial z^j} \right|_{0}
\int_0^h \mathrm{d}z \, \frac{z^{i+j}}{i!j!} \frac{\tilde{\chi}}{1 + N(z) \tilde{\chi}}
- \mu_0 \sum_{n=2}^{\infty}  \left. \frac{\partial^n H_a}{\partial z^n} \right|_{0} 
\frac{ M(H_a(0))}{1 + N(h) \tilde{\chi}} \frac{h^{n+1}}{(n+1)!} \\
- \mu_0 M(H_a(0)) \sum_{n=1}^{\infty}  \left. \frac{\partial^n H_a}{\partial z^n} \right|_{0}
\int_0^h \mathrm{d}z \, \frac{\tilde{\chi}}{(1 + N(z) \tilde{\chi})^2} \frac{z^{n+1}}{(n+1)!} \frac{\partial N}{\partial z} \qquad
\label{master_equation}
\end{multline}

The first term is a volume force due to gravity and an additional force due to the uniform contribution of the gradient. The second term is the usual surface tension term. The third term is the origin of the normal-field instability. Note that it has been separated such that it is exactly in the same shape as for a ferrofluid in presence of an uniform field $H_a(0)$, and plays the exact same role if the demagnetization is assumed to be uniform inside the fluid (weak gradient or thin distribution limits).

All other terms are non-linear terms. The fourth and the fifth may not be negligible if the gradient is strong. The sixth term only contains higher order magnetic field derivatives multiplying $h^n$, $n \geq 3$ terms and can usually be neglected, but one can easily include some of them if needed in order to improve the accuracy of a numerical resolution.

\bibliographystyle{jfm}
\bibliography{mybib.bib}

\end{document}